\documentclass[11pt, a4paper]{article}
\usepackage{jheppub}
\usepackage{graphicx}
\usepackage{bm}

\renewcommand{\d}{\delta}
\newcommand{\D}{\Delta}

\newcommand{\g}{\gamma}

\newcommand{\m}{\mu}

\renewcommand{\S}{\Sigma}

\renewcommand{\t}{\tau}

\newcommand{\ol}[1]{\overline{#1}}

\newcommand{\PCAC}{\text{PCAC}}

\newcommand{\ChPT}{\text{ChPT}}

\newcommand{\LO}{\text{LO}}
\newcommand{\NLO}{\text{NLO}}
\newcommand{\SLO}{\text{SLO}}
\newcommand{\NSLO}{\text{NSLO}}
\newcommand{\Loop}{{1\text{loop}}}

\newcommand{\p}{\partial}
\newcommand{\lag}{\mathcal L}

\newcommand{\tchi}{\tilde \chi}
\newcommand{\hata}{\hat A}
\newcommand{\bpsi}{\ol{\psi}}

\newcommand{\tilm}{\tilde m}
\newcommand{\tV}{\tilde V}
\newcommand{\tX}{\tilde X}

\newcommand{\vev}[1]{{\langle #1 \rangle}}
\renewcommand{\eqref}[1]{(\ref{#1})}

\begin{document}
\title{Chiral perturbation theory for twisted mass QCD at small quark mass}
\author[a]{Satoru Ueda}
\author[b,c]{and Sinya Aoki}
\affiliation[a]{KEK Theory Center, High Energy Accelerator
Research Organization (KEK),\\
 Tsukuba 305-0801, Japan}
\affiliation[b]{Graduate School of Pure and Applied Sciences,
University of Tsukuba,\\
 Tsukuba, Ibaraki 305-8571, Japan}
\affiliation[c]{Center for Computational Sciences, University
of Tsukuba,\\
 Tsukuba, Ibaraki 305-8577, Japan}
\emailAdd{sueda@post.kek.jp}
\emailAdd{saoki@het.ph.tsukuba.ac.jp}
\abstract{
We study the lattice cutoff ($a$) and quark mass dependences of pion masses
and decay constants in the $N_f = 2$ twisted mass QCD,
using the Wilson chiral perturbation theory to the next leading order (NLO).
In order to investigate the region near zero quark mass,
we introduce the power counting scheme where $O(a^2, am)$ terms
are included in the tree level effective Lagrangian.
At the NLO of  this power counting scheme,
we calculate the charged pion mass and decay constant as
a function of the lattice cutoff as well as the twisted quark mass
at the maximal twist.
In this paper, we adopt two different definitions for the maximal twist.
We confirm that  the difference between the two appears as
the $O(a^2)$ effects so that the automatic $O(a)$ improvement
is realized for both definitions.}

\maketitle
%
%
\section{introduction}

The twisted mass lattice QCD (tmlQCD)
\cite{Frezzotti:1999vv,Frezzotti:2000nk,Frezzotti:2001ea}
has several advantages for numerical simulations,
one of which is the automatic $O(a)$ improvement
at the maximal twist \cite{Frezzotti:2001ea,Aoki:2004ta,Aoki:2006nv}.
The tmlQCD becomes free from $O(a)$ lattice spacing errors,
by simply setting the twist angle to its maximum value $\omega = \pi/2$.

This automatic $O(a)$ improvement of the tmlQCD has been investigated in
quench simulations by the XLF Collaboration
\cite{Jansen:2003ir,Bietenholz:2004wv,Jansen:2005gf,Jansen:2005kk}
and Abedel-Rehime et al.\cite{AbdelRehim:2004gx,AbdelRehim:2005gz},
while the unexpected first order phase transition has been found in
full QCD simulations by DESY group
\cite{Farchioni:2004us,Farchioni:2004ma,Farchioni:2005tu}.
Recently the European Twisted mass Collaboration (ETMC) has started
the large-scale full QCD simulations at the maximal twist with
$N_f = 2$ \cite{Alexandrou:2008tn,Baron:2009wt} and
$N_f = 2 + 1 + 1$ \cite{Baron:2008xa,Baron:2010th}.

The chiral perturbation theory (ChPT)
\cite{Gasser:1983ky,Gasser:1983kx,Gasser:1983yg,Gasser:1984gg},
which is a low energy effective theory of QCD describing the dynamics
of Nambu-Goldstone(NG) bosons, plays an important role to extrapolate
physical observables such as NG boson masses and decay constants calculated
in lattice QCD simulations at heavier quark masses
to the physical quark mass point.
Furthermore, not only the quark mass dependence but also
the scaling violation for the Wilson quark action are described by
the Wilson ChPT (WChPT), which includes effects of non-zero
lattice spacing $a$ \cite{Sharpe:1998xm,Sharpe:2004ps,Aoki:2004pq}.
Since $O(a)$ contribution can be absorbed into the quark mass term $m$,
an inclusion of the lattice spacing effect in the ChPT is rather non-trivial, so that
$O(a^2)$ terms dominate in the small quark mass region.
For QCD with the ordinary Wilson quark, several physical observables
have been calculated in the WChPT at the next leading order (NLO),
while only the leading order (LO) results exist in the WChPT for the tmlQCD%
\footnote{There exists the NLO ChPT calculations without $O(a^2, am)$ terms
for $N_f = 2$ \cite{Munster:2003ba,Scorzato:2004da,Sharpe:2004ny},
$N_f = 3$ \cite{Munster:2003ba,Sharpe:2004ny}, and
$N_f = 2 + 1 + 1$ \cite{Munster:2011gh}.}
\footnote{In ref.~\cite{Bar:2010jk},  NLO chiral log terms have been calculated
with a similar power counting scheme, but in this paper we calculate and
consider the renormalization at NLO in detail and the general case.}.
In this paper, we therefore present  results of pion masses
and decay constants in the WChPT at the NLO,
including $O(a^2, am)$ terms in the LO Lagrangian.

In Sec.~\ref{sec:leading_order}, we consider the power counting scheme
in detail and give results at LO.
In Sec.~\ref{sec:next_leading_order}, we calculate the one loop contribution
for pion masses and decay constants, and show that divergences of
these quantities at one loop can be renormalized by the NLO counter terms.
We present the pion masses and decay constants at NLO arbitrary value
of the twist angle.
In Sec.~\ref{sec:maximal_twist}, we consider the automatic $O(a)$ improvement
at the maximal twist, using the NLO calculation.
We summarize our paper in Sec.~\ref{sec:Conclustion}.
Details of NLO calculations are given in appendix.~\ref{sec:app_detail_calc}.

%
%
\section{Analysis at "leading order"}
\label{sec:leading_order}

\subsection{Power counting and Lagrangian}
\label{sec:powercounting}

In the ordinary continuum ChPT~\cite{Gasser:1983ky,Gasser:1983kx,Gasser:1983yg, Gasser:1984gg}, the quark mass $m$, equivalently the meson momentum $p^2$, is considered as the expansion parameter, so that
$O(M)$ terms consist of the LO Lagrangian where $M=m$ or $p^2$.
At the NLO order, local counter terms of $O(M^2)$  cancel
the divergences of one loop contributions generated by the LO terms.

 In addition to $M$, the lattice spacing $a$ appears as the expansion parameter in the chiral perturbation theory for lattice QCD. Due to the explicit breaking of the chiral symmetry in the Wilson quark action, the lattice spacing effects start at $O(a)$. Therefore it is natural to treat $O(a)$ term as the LO contribution such that $a\sim M$.  In the WChPT, this LO $O(a)$ term can be absorbed into the mass term by
\begin{equation}
 m \to \tilde m = m + O(a),
\end{equation}
so that no extra contributions  to the continuum ChPT appear  in the WChPT at this order.
This power counting, however, is not correct in the region where $\tilde m$ is small, since the NG boson mass $m_\pi$ at the LO is given by
\begin{eqnarray}
m_\pi^2 &=& 2 B \tilde m,
\end{eqnarray}
which becomes tachyon for negative $\tilde m$%
\footnote{In the continuum ChPT,  no tachyon appears for all $m$,
since $\tilde m\rightarrow \vert m\vert$ in the mass formula,
thanks to the chiral symmetry, which is absent in the WChPT.}.
One has to add $O(a^2)$ terms to the LO Lagrangian for the stability of the vacuum to avoid the appearance of the tachyon\cite{Sharpe:1998xm,Aoki:2004pq}.
This LO Lagrangian is still insufficient due to the following reason.
While 1-loop contributions from this LO Lagrangian generate $O(\tilde M^2, \tilde M a^2,a^4)$ terms where $\tilde M=\tilde m$ or $p^2$, terms odd in $a$ such as $O(\tilde M a)$ are never generated from 1-loop contributions.  Hereafter we ignore $O(a^4)$ terms since they are small in present lattice QCD simulations. 

From the above consideration, a physical observable $X$ in general depends on both quark mass $\tilde m$ and $a$ as
\begin{eqnarray}
X(\tilde m,a) &=& \tilde m X_0(\tilde m) + \tilde m a X_1(\tilde m) + a^2 X_2(\tilde m) +O(a^3),
\end{eqnarray}
where we ignore small $a^3$ or higher contributions. 
We therefore set up the WChPT to calculate $X_0(\tilde m)$, $X_1(\tilde m)$ and $X_2(\tilde m)$ order by order in $\tilde m$. For this purpose, in addition to leading order $O(\tilde M)$ terms, 
we consider $O(a^2, \tilde M a)$ terms as the tree level Lagrangian.   We call  $O(a^2, \tilde M a)$ terms the sub-leading order (SLO) terms to distinguish them from LO terms.
 The one-loop contributions generate an additional factor of $\tilde M$ to the tree level Lagrangian, LO plus SLO. Divergences from these one-loop contributions must be canceled by NLO $O(\tilde M^2)$ and NSLO $O(a^2\tilde M, a\tilde M^2)$ terms.
We stress again that higher order contributions such as $a^2\tilde M^2$ are neglected here.

For the twisted mass fermion~\cite{Frezzotti:1999vv,Frezzotti:2000nk},
there are two mass parameters, untwisted quark mass $\tilde m$ and twisted quark
mass $\mu$, which are denoted as $\bm{m} = {\tilde m, \mu}$.
Therefore $\tilde M$ in this case represents $\tilde m$, $\mu$ or $p^2$.
In table \ref{t-power_counting}, we summarize our power counting scheme.
\begin{table}[tb]
\begin{center}
\begin{tabular}{|c|c|l|c|l|}
 \hline
 tree level  & LO & $O(\tilde M)$
 & SLO & $O(a^2, a \tilde M)$\\
 \hline
 one loop & NLO & $O(\tilde M^2)$ 
  & NSLO & $O( a^2\tilde M, a\tilde M^2)$\\
\hline
\end{tabular}
\caption{Power counting scheme in this paper. We treat $O(\tilde M)$ (LO) and
$O(a^2, a\tilde M)$ (SLO) as the tree level Lagrangian.
We introduce NLO  and NSLO terms as local counter terms to cancel divergence of one loop contributions.}
\label{t-power_counting}
\end{center}
\end{table}

\subsection{Tree-level Lagrangian}

The tree-level effective Lagrangian for $N_f=2$ tmlQCD,
which include LO and SLO, is given by
\begin{align}
\lag_{\LO} &= \frac{f_0^2}{4}\vev{D_\m\S D_\m\S^\dag}
- \frac{f_0^2}{4}\vev{\S \chi^\dag + \chi \S^\dag}
- \frac{f_0^2}{4}\vev{\S \hata^\dag + \hata \S^\dag}
\label{eq:LO_action}\\
\lag_{\SLO} &= W_{45} \vev{D_\m\S D_\m\S^\dag}
\vev{(\S-\S_0) \hata^\dag + \hata (\S-\S_0)^\dag}\notag\\
&- W_{68}\vev{\S \chi^\dag + \chi \S^\dag}
\vev{\S \hata^\dag + \hata \S^\dag}
- W_{68}^\prime\vev{\S \hata^\dag + \hata \S^\dag}^2,
\label{eq:tree-level_tmWChPT_action}
\end{align}
where $D_\m\S$ is covariant derivative with the left and
right source current $l_\mu$ and $r_\mu$ defined by
\begin{align}
 D_\mu\Sigma &= \partial_\mu\Sigma - il_\mu\Sigma + i\Sigma r_\mu,\\
 D_\mu\Sigma^\dag &= \partial_\mu\Sigma^\dag + i\Sigma^\dag l_\mu
- ir_\mu \Sigma.
\end{align}
After the construction of the Lagrangian,
spurion fields $\chi$ and $\hata$ should be set to
\begin{align}
\chi &\to 2B_0 M = 2B_0(m + i\t^3\m),& \hata &\to 2W_0 a.
\label{tmWChPT_spurion_fields}
\end{align}
The coefficients $f_0, B_0, W_{0, 45, 68}$ and $W'_{68}$
in eqs. (\ref{eq:LO_action}), (\ref{eq:tree-level_tmWChPT_action}) and
(\ref{tmWChPT_spurion_fields}) are the low energy constants,
and their dimension are $[f_0] = [B_0] = 1, [W_0] = 3$,
and $[W_{45, 68}] = [W_{68}] = 0$.
The $SU(2)$ matrix fields $\S(x)$ for NG bosons is defined by
\begin{align}
\S &= \S_0^{1/2} \S_{\text{ph}} \S_0^{1/2},\label{eq:def_WChPT_vacuum}\\
\S_\text{ph} &= \exp\left[i \frac{\pi_a(x)\t^a}{f_0}\right],
\label{eq:def_WChPT_field}
\end{align}
where $\S_0$ is a vacuum expectation value of $\S(x)$,
$\t^a$ is the Pauli matrices and $\pi_a(x)$ is the pseudo scalar NG field.

As already mentioned, the $O(a)$ term, $\vev{\S \hata^\dag + \hata \S^\dag}$
can be absorbed to
$O(\bm{m})$ term, $\vev{\S \chi^\dag + \chi \S^\dag}$ by the replacement that
\begin{equation}
2B_0 \tilde m = 2B_0 m + 2W_0 a.
\end{equation}
Replacing the mass parameter $m$ with shifted mass $\tilm$,
we obtain
\begin{align}
\lag_{\LO + \SLO} &= \frac{f_0^2}{4}
\left[1 + \frac{c_0a}{4} \vev{(\S-\S_0) + (\S-\S_0)^\dag}\right]
\vev{D_\m\S D_\m\S^\dag}\notag\\
& -f_0^2\biggl[\frac{2B_0\tilm}{4}\vev{\S+\S^\dag}
- \frac{2B_0\m}{4}\vev{i(\S-\S^\dag)\t^3}\notag\\
&\quad + \frac{\tilde c_2a 2B_0 \tilm - c_2a^2}{16}\vev{\S+\S^\dag}^2
- \frac{\tilde c_2a 2B_0 \m}{16} \vev{\S+\S^\dag}\vev{i(\S-\S^\dag)\t^3}
\biggr],
\label{eq:lagrangian_tree}
\end{align}
where the coefficients $c$'s are defined by
\begin{align}
c_0 &= 32 W_{45} \frac{W_0}{f_0^2},&
c_2 &= - 64(W'_{68} - W_{68}) \frac{W_0^2}{f_0^2},&
\tilde c_2 &= 32 W_{68}\frac{W_0}{f_0^2}.
\end{align}

\subsection{Gap equation and pseudo scalar meson mass}

We first determine the vacuum in the Lagrangian (\ref{eq:lagrangian_tree}).
Parametrizing the vacuum expectation value of $\S$ as
\begin{equation}
\S_0 = \exp[i\phi \t^3] = \cos\phi + i \t^3 \sin\phi,
\end{equation}
the vacuum energy becomes
\begin{equation}
V(\phi) = - f^2_0 [2B_0\tilm \cos\phi + 2B_0\m \sin\phi
- (c_2a^2 - \tilde c_2 a 2B_0\tilm) \cos^2\phi
+ \tilde c_2a 2B_0\m\cos\phi\sin\phi].
\label{vacuum_energy}
\end{equation}
The vacuum expectation value corresponds to the minimal point of
this vacuum energy (\ref{vacuum_energy}) $\phi_0$,
which is determined by solving the  gap equation,
\begin{equation}
2B_0 \tilm \sin\phi_0 - (c_2 a^2 - \tilde c_2 a 2B_0\tilm)\sin 2\phi_0
= 2B_0\m \cos\phi_0 + \tilde c_2a 2B_0\m \cos 2\phi_0.
\label{eq:WChPT_gap_eq}
\end{equation}

We next expand the Lagrangian in terms of the component fields $\pi$ as
\begin{equation}
\lag_{\LO + \SLO} = \lag^{(2)}_{\LO + \SLO} + \lag^{(3)}_{\LO + \SLO} + \cdots,
\end{equation}
where $\lag_{\LO + \SLO}^{(n)}$ represents the $O(\pi^n)$ terms
in the tree-level Lagrangian.
With the components fields $\pi_a$,
the $O(\pi^2)$ Lagrangian $\lag_{\LO + \SLO}^{(2)}$ is give by
\begin{equation}
\lag_{\LO + \SLO}^{(2)} = \frac{1}{2} (\p_\m \pi_a)^2
+ \frac{m_{\pi,a}^2}{2} ( \pi_a)^2.
\end{equation}
where the tree-level pion mass $m_{\pi,a}^2$ is written as
\begin{align}
m_{\pi,a}^2 &= \begin{cases}
(m_\pi^\pm)^2 = m_\pi^2 & \qquad (a = 1,2)\\
(m_\pi^0)^2 = m_\pi^2 + \D m_\pi^2 & \qquad (a = 3)
\end{cases},\\
m_\pi^2 &= 2B_0 m' - 2c_2a^2\cos^2\phi_0 + 2\tilde c_2a(2B_0m')\cos\phi_0,\\
\D m_\pi^2 &= 2c_2a^2 \sin^2\phi_0 + 2\tilde c_2 a (2B_0\m')\sin\phi_0 .
\end{align}
Here $m_\pi \equiv m_\pi^\pm$ ($m_\pi^0$) denotes
the charged (neutral) pion mass, and the short-handed notation
for mass parameter is given by
\begin{equation}
\left(\begin{array}{c}
m' \\ \m'
\end{array}\right) = \left(\begin{array}{cc}
\cos\phi_0 & \sin\phi_0 \\ - \sin\phi_0 & \cos\phi_0
\end{array}\right)\left(\begin{array}{c}
\tilde m \\ \m
\end{array}\right).
\end{equation}
Using this notation, the gap equation (\ref{eq:WChPT_gap_eq}) is written as
\begin{equation}
 2B_0 \m' = - c_2a^2 \sin2\phi + \tilde c_2a(2B_0m' \sin\phi - 2B_0\m'\cos\phi).
\label{eq:WChPT_gap_eq2}
\end{equation}
Using this recursively, $a\m'$ terms is found to be be of higher order, so that
we can replace the $\m'$ term with $a^2, am'$ terms.

We now discuss the lattice cutoff $a$ as well as quark masses $\tilde m, \m$
dependences of vacuum condensation $\cos\phi_0$ and pion masses.
Let us consider the case that the coefficient $\tilde c_2 = 0$.
In this case, there are two possible phase diagrams\cite{Sharpe:2004ps},
depending on the sign of a coefficient $c_2$.
Solving the gap equation (\ref{eq:WChPT_gap_eq})
and minimizing the potential energy (\ref{vacuum_energy}),
we obtain the dependence of $\cos\phi_0$ on the untwisted quark mass $\tilde{m}$.
\begin{figure}[htbp]
\begin{center}
\includegraphics[width=.5\textwidth]{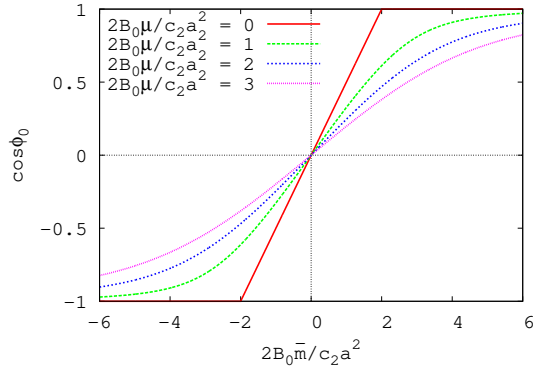}
\caption{Vacuum angle $\cos\phi_0$ as a function of $\tilde m$
for $c_2>0$ and $2B_0\m/c_2 a^2 = 0,1,2,3$.}
\label{fig:c2p_m_cphi_tc20}
\end{center}
\end{figure}
\begin{figure}[htbp]
\begin{minipage}{0.5\textwidth}
\begin{center}
\includegraphics[width=0.8\textwidth]{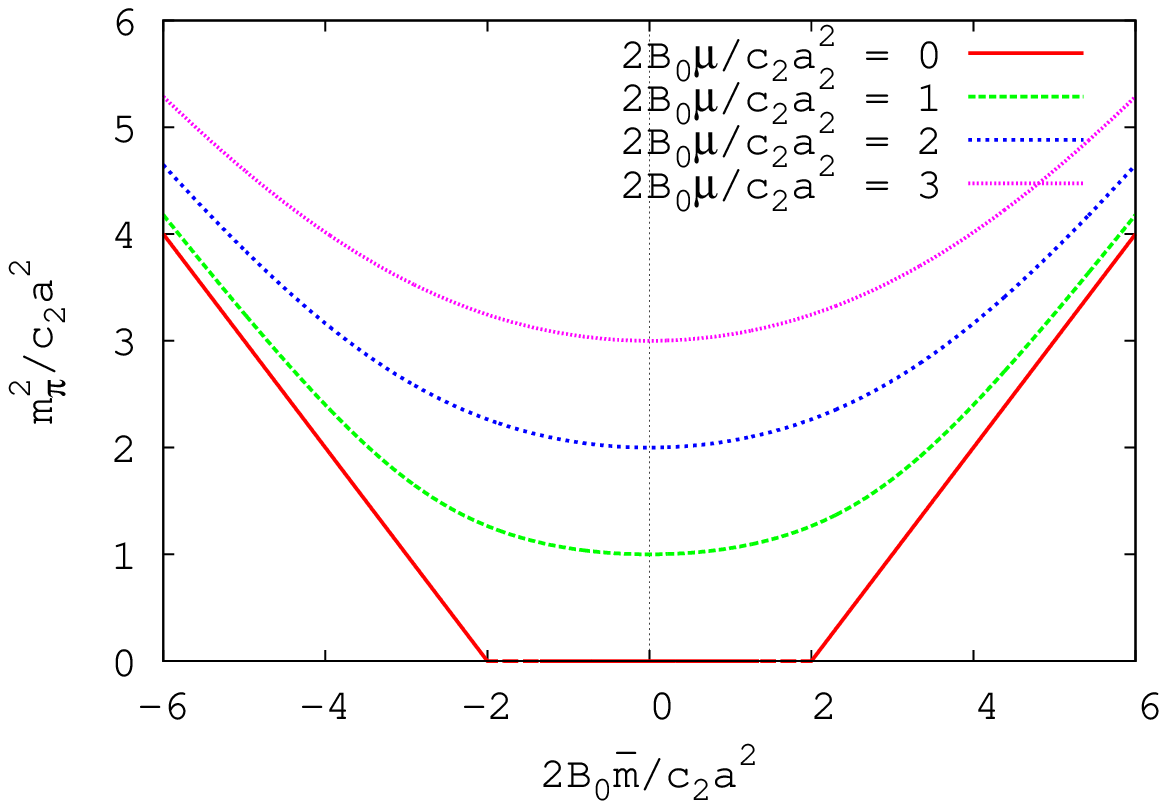}\\
(a) Charged pion mass $(m_\pi^\pm)^2$.
\end{center}
\end{minipage}
\begin{minipage}{0.5\textwidth}
\begin{center}
\includegraphics[width=0.8\textwidth]{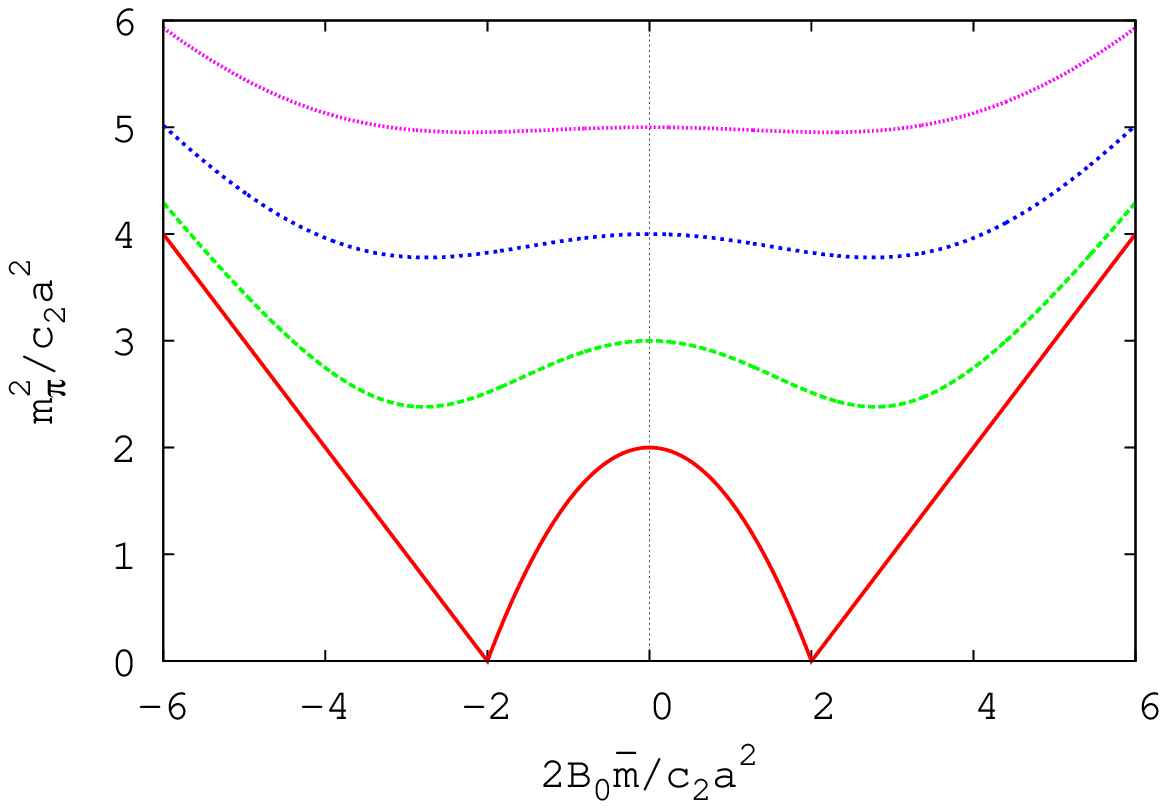}\\
(b) Neutral pion mass $(m_\pi^0)^2$.
\end{center}
\end{minipage}
\caption{Pion masses as a function of $\tilde m$ for $c_2 > 0$
and $2B_0\m/c_2 a^2 = 0,1,2,3$.}
\label{fig:2nd_pion_mass}
\end{figure}
For $c_2 > 0$, figure~\ref{fig:c2p_m_cphi_tc20} shows the form of
$\cos\phi_0$ as a function of $2B_0 \tilde m/c_2 a^2$
at $2B_0\m/c_2a^2 = 0, 1, 2,$ and $3$.
Corresponding pion masses are shown in figure \ref{fig:2nd_pion_mass}.
In the untwisted theory ($\m = 0$),
there are second order transitions at $2B_0\tilde m / c_2a^2 = \pm 2$,
as shown by the kinks in figure.~\ref{fig:c2p_m_cphi_tc20}
and by the vanishing pion masses in figure~\ref{fig:2nd_pion_mass}.
The parity-flavor breaking phase, defined by
the condition that $\cos\phi_0 \neq \pm 1$,
lies between these two second order phase transition points,
and two Nambu-Goldstone bosons associated with the flavor breaking
appear in this phase\cite{Aoki:1983qi,Aoki:1986xr,Aoki:1995ft}.
Once $2B_0\m$ becomes non-zero, however, the transition turns into a crossover,
and pion masses always stay non-zero due to
the explicit flavor breaking by non-zero $\mu$.
For $c_2 > 0$, the charged pion is heavier than the neutral pion,
due to the $O(a^2)$ effect.
If we change the value of the twisted mass continuously
at fixed $\vert 2B_0\tilde m / c_2a^2\vert < 2$,
there appears the first order phase transition
while crossing the parity-breaking phase at $\mu=0$.
\begin{figure}[btp]
\begin{minipage}{0.5\textwidth}
\begin{center}
\includegraphics[width=0.8\textwidth]{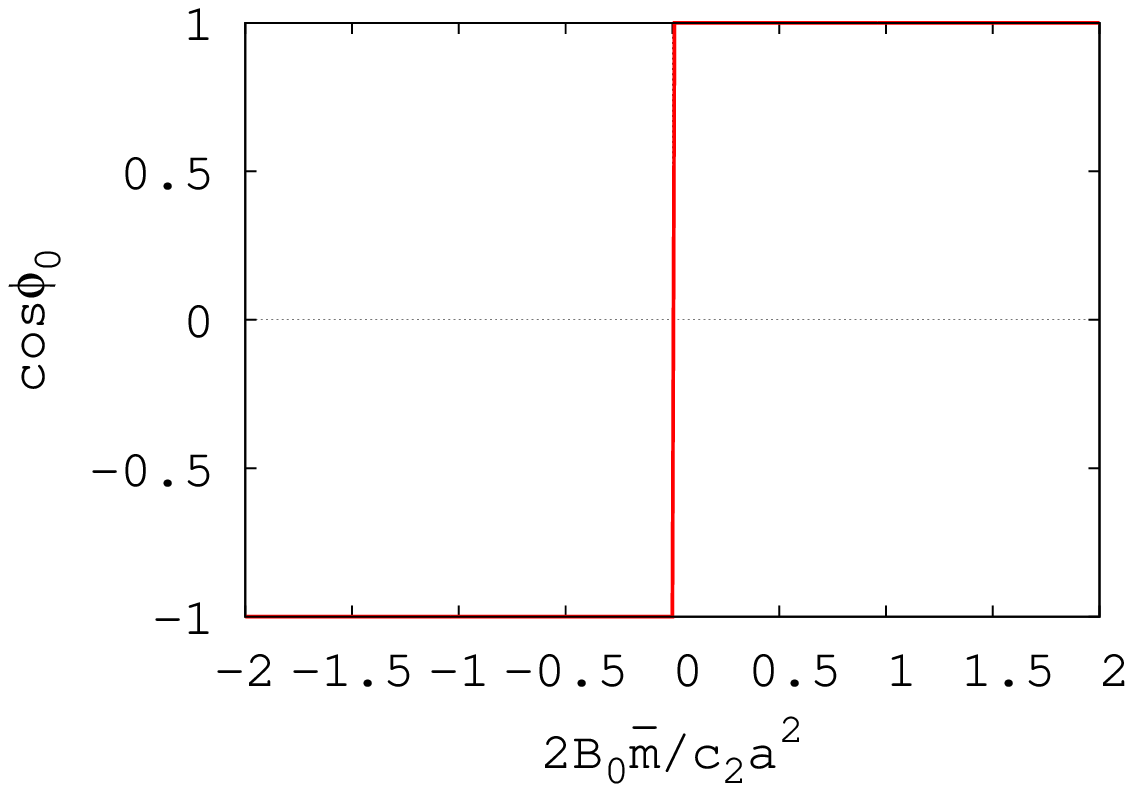}\\
(a) Vacuum angle at $2B_0\m = 0$
\end{center}
\end{minipage}
\begin{minipage}{0.5\textwidth}
\begin{center}
\includegraphics[width=0.8\textwidth]{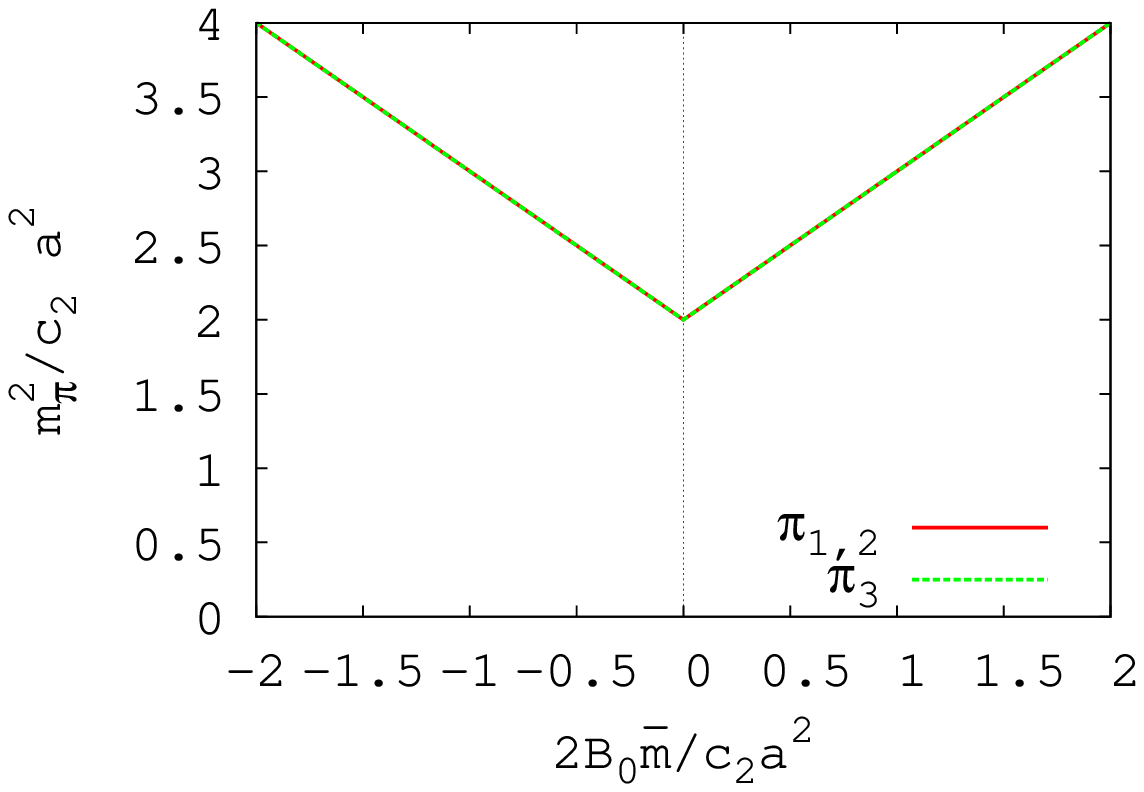}\\
(b) Pion masses at $2B_0\m = 0$
\end{center}
\end{minipage}
\begin{minipage}{0.5\textwidth}
\begin{center}
\includegraphics[width=0.8\textwidth]{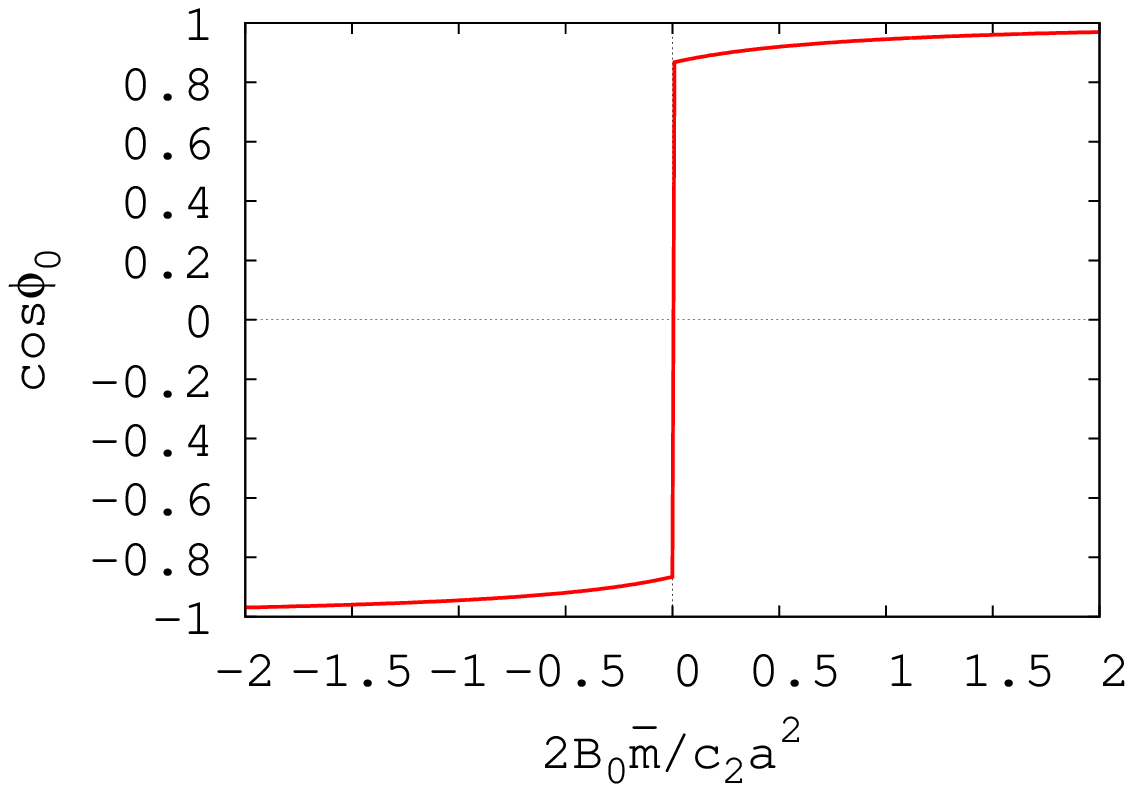}\\
(c) Vacuum angle at $2B_0\m = c_2a^2$
\end{center}
\end{minipage}
\begin{minipage}{0.5\textwidth}
\begin{center}
\includegraphics[width=0.8\textwidth]{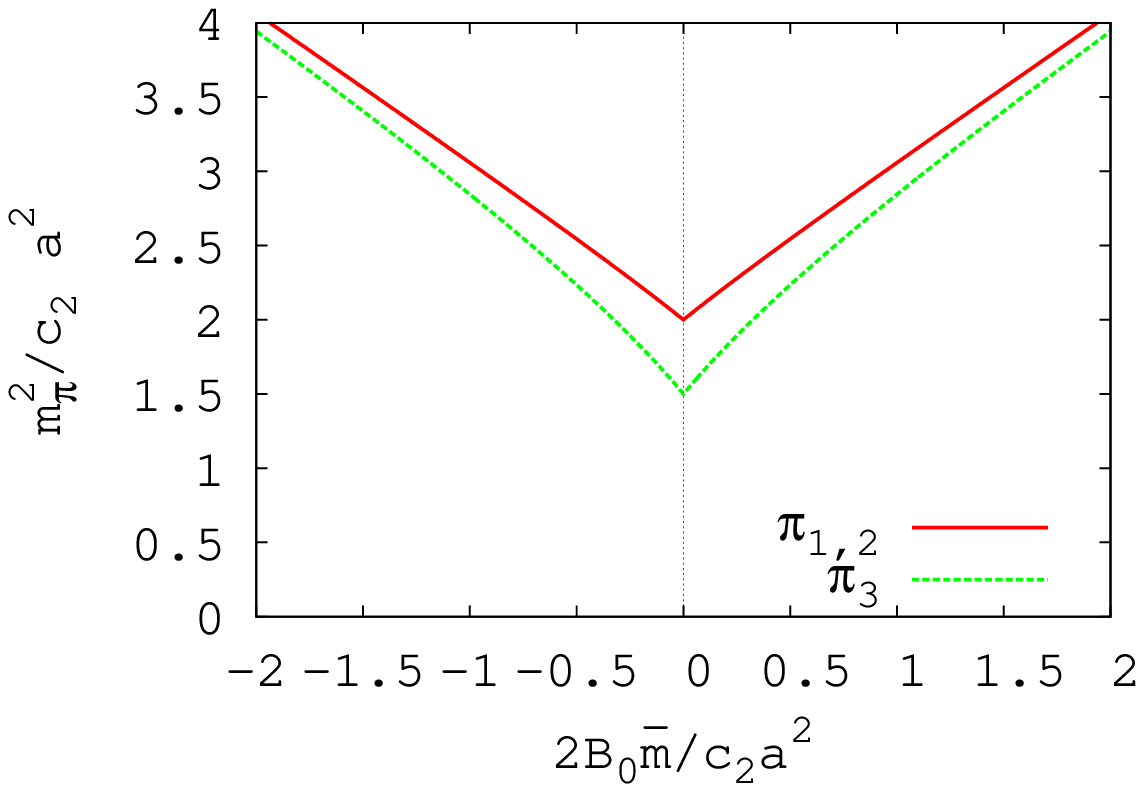}\\
(d) Pion masses at $2B_0\m = c_2a^2$
\end{center}
\end{minipage}
\begin{minipage}{0.5\textwidth}
\begin{center}
\includegraphics[width=0.8\textwidth]{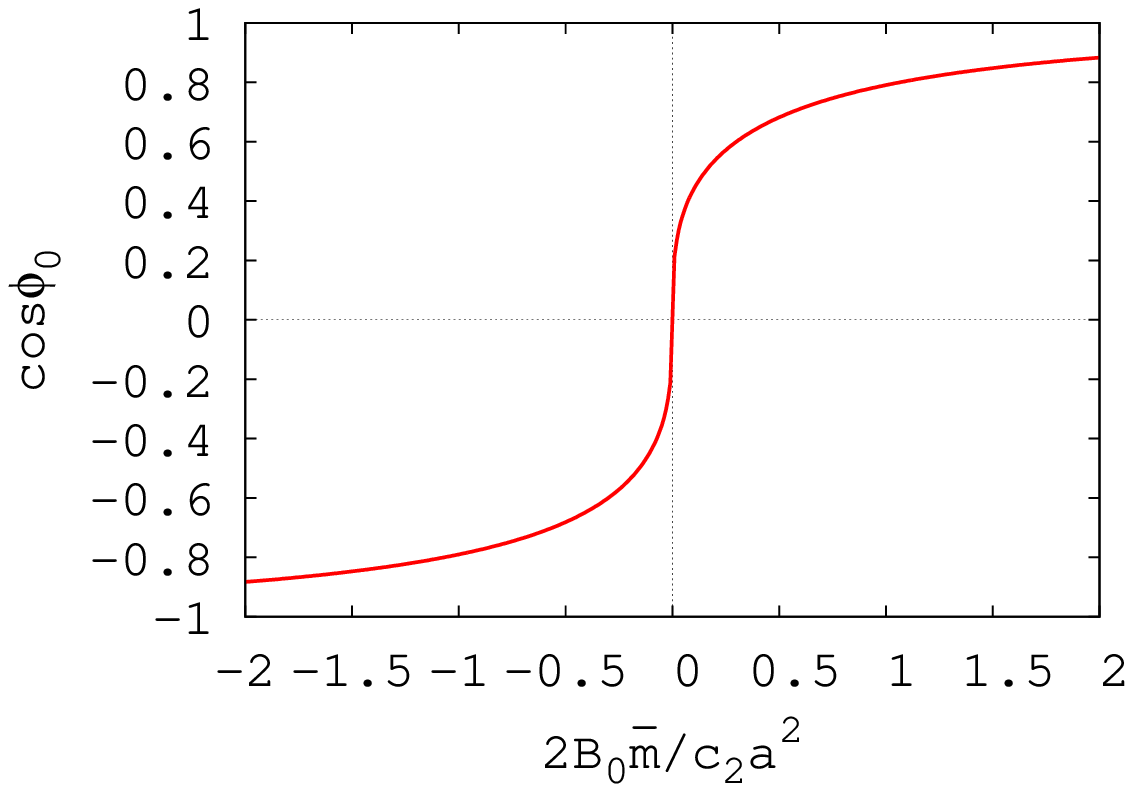}\\
(e) Vacuum angle at $2B_0\m = 2c_2a^2$
\end{center}
\end{minipage}
\begin{minipage}{0.5\textwidth}
\begin{center}
\includegraphics[width=0.8\textwidth]{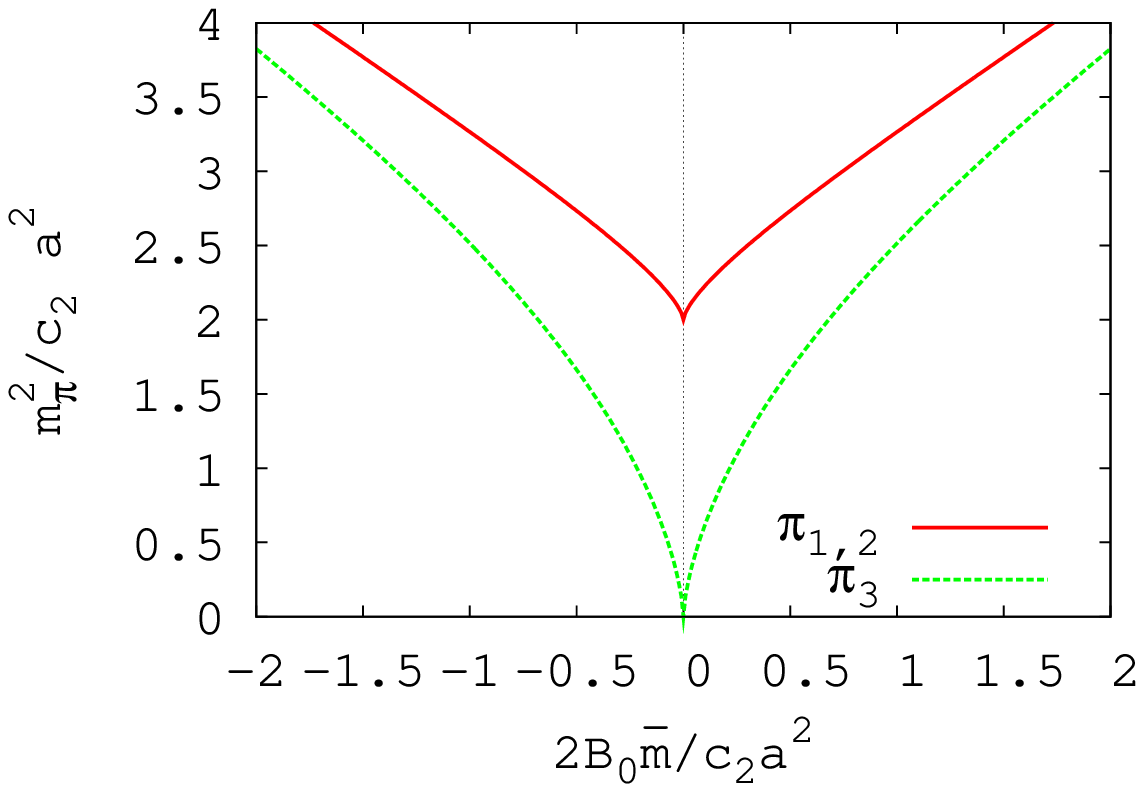}\\
(f) Pion masses at $2B_0\m = 2c_2a^2$
\end{center}
\end{minipage}
\begin{minipage}{0.5\textwidth}
\begin{center}
\includegraphics[width=0.8\textwidth]{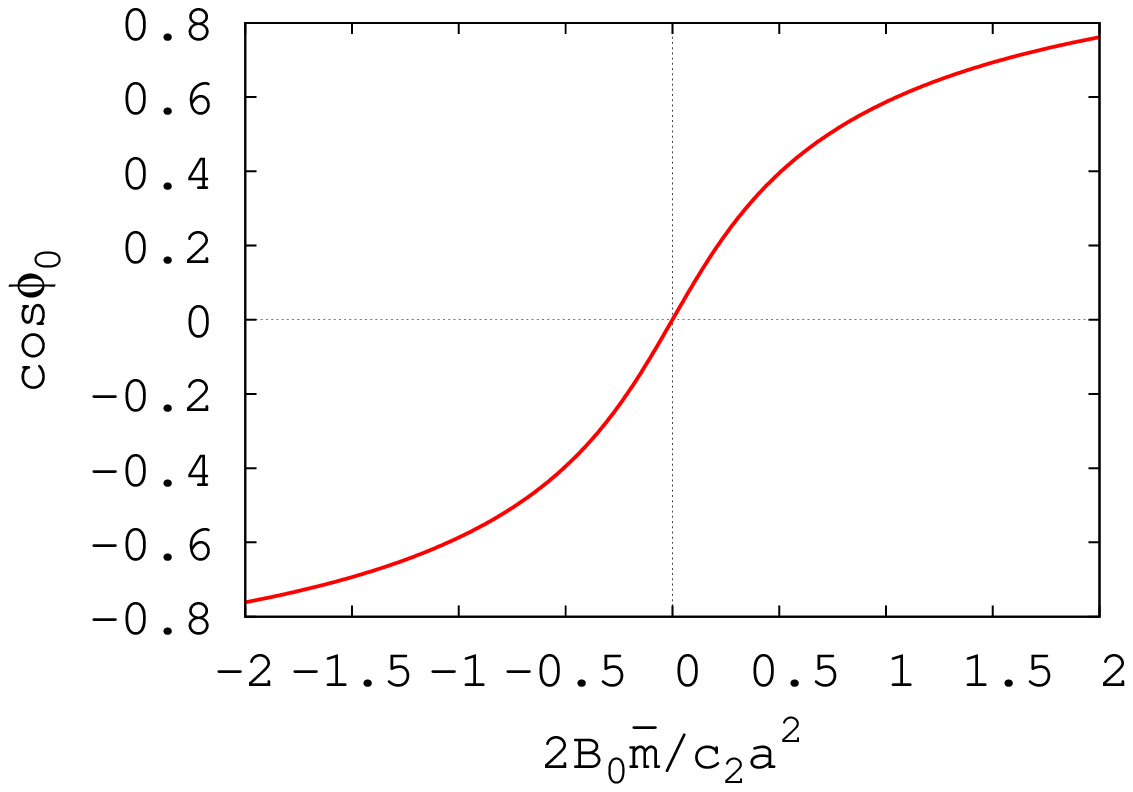}\\
(g) Vacuum angle at $2B_0\m = 3c_2a^2$
\end{center}
\end{minipage}
\begin{minipage}{0.5\textwidth}
\begin{center}
\includegraphics[width=0.8\textwidth]{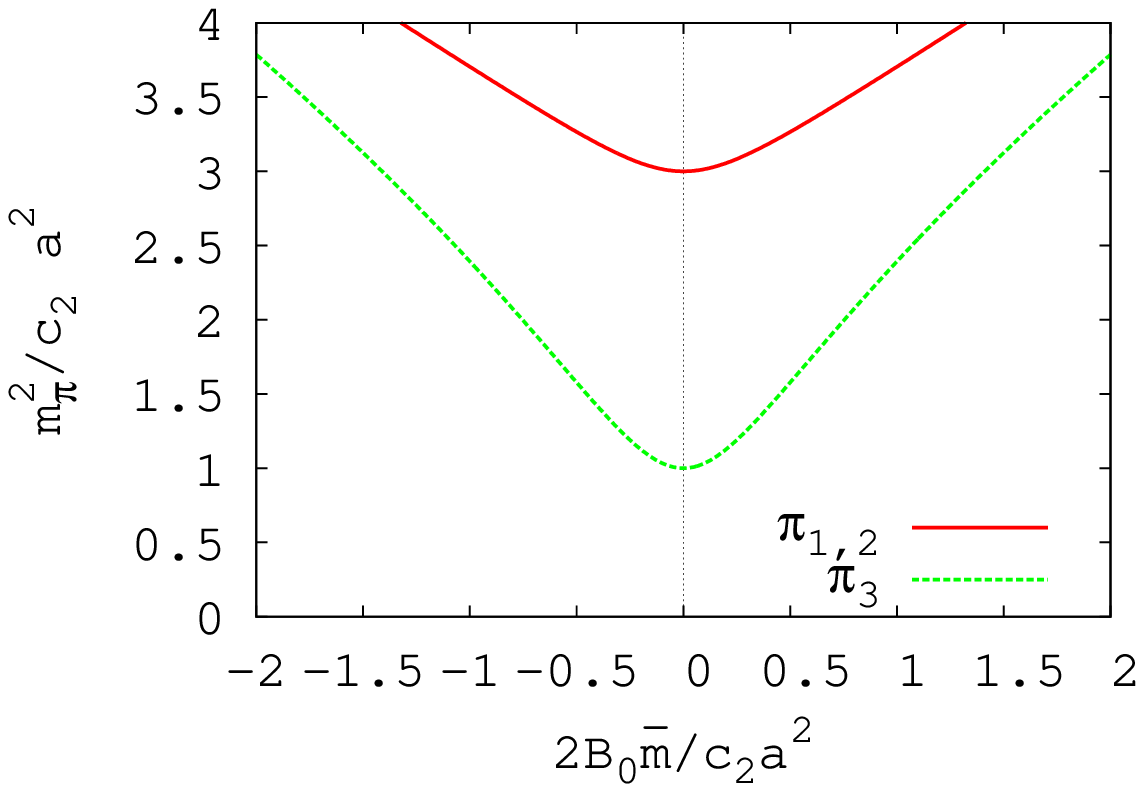}\\
(h) Pion masses at $2B_0\m = 3c_2a^2$
\end{center}
\end{minipage}
\caption{Vacuum angle $\cos\phi_0$ and pion masses
as a function of $\tilde m$ for $c_2<0$ at $2B_0\m/c_2 a^2 = 0,1,2,3$.}
\label{fig:1st_phse_mu0}
\end{figure}

For $c_2 < 0$, figure~\ref{fig:1st_phse_mu0}
shows $\cos\phi_0$ and pion masses as a function of $2B_0\tilde m/c_2a^2$
at the several fixed values of $2B_0\m$.
Figures~\ref{fig:1st_phse_mu0}-a and \ref{fig:1st_phse_mu0}-b
show the results for untwisted theory at $\m = 0$.
The condensate $\cos\phi_0$ jumps from $\S_0 = 1$ for $2B_0\tilde m > 0$
to $\S_0 = -1$ for $2B_0\tilde m < 0$.
This is the first order transition without the flavor breaking,
so that all pions remain massive and degenerate.
The effect of non-zero twisted mass $\m$ generates a non-zero value of $\t^3$ component, $\sin\phi_0=\sqrt{1-\cos^2\phi}$.
There still remains the first order phase transition
at which $\sin \phi_0$ flips sign between $\pm (1 - \vert 2B_0\m /2c_2a^2\vert)$. The first order phase transition disappears at $\vert 2B_0\m /2c_2a^2\vert = 2$ and it turns into
a cross-over at $\vert 2B_0\m /2c_2a^2\vert > 2$.
Due to the explicit flavor breaking,
the neutral pion($\pi^3$) is lighter than charged pions($\pi^\pm$).

\begin{figure}[btp]
\begin{minipage}{0.5\textwidth}
\begin{center}
\includegraphics[width=0.8\textwidth]{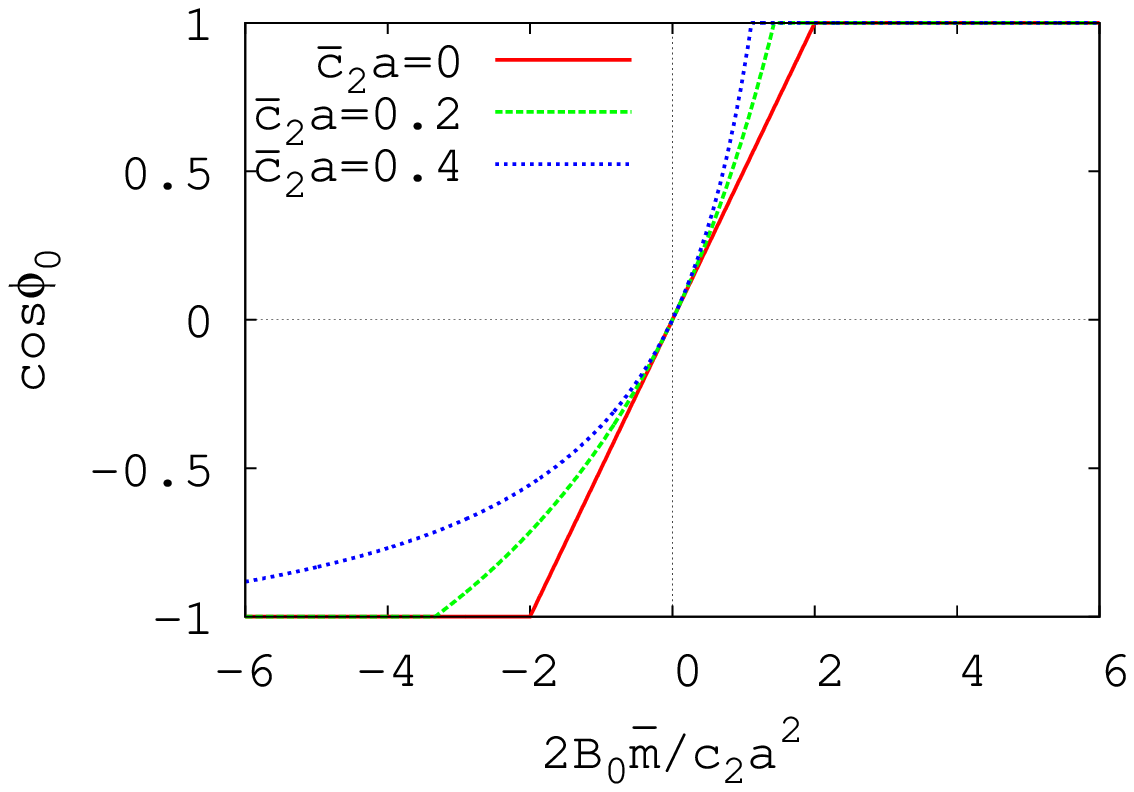}\\
(a) Vacuum angle for $c_2 > 0$
\end{center}
\end{minipage}
\begin{minipage}{0.5\textwidth}
\begin{center}
\includegraphics[width=0.8\textwidth]{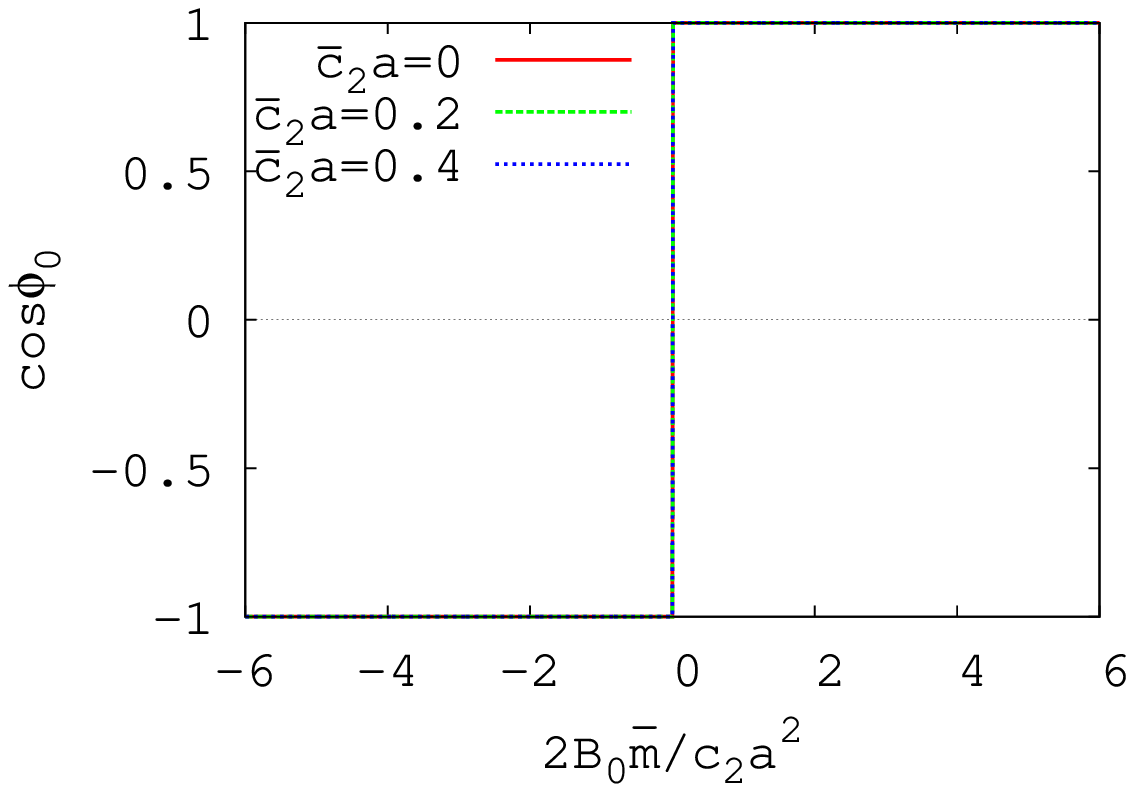}\\
(b) Vacuum angle for $c_2 < 0$
\end{center}
\end{minipage}
\begin{minipage}{0.5\textwidth}
\begin{center}
\includegraphics[width=0.8\textwidth]{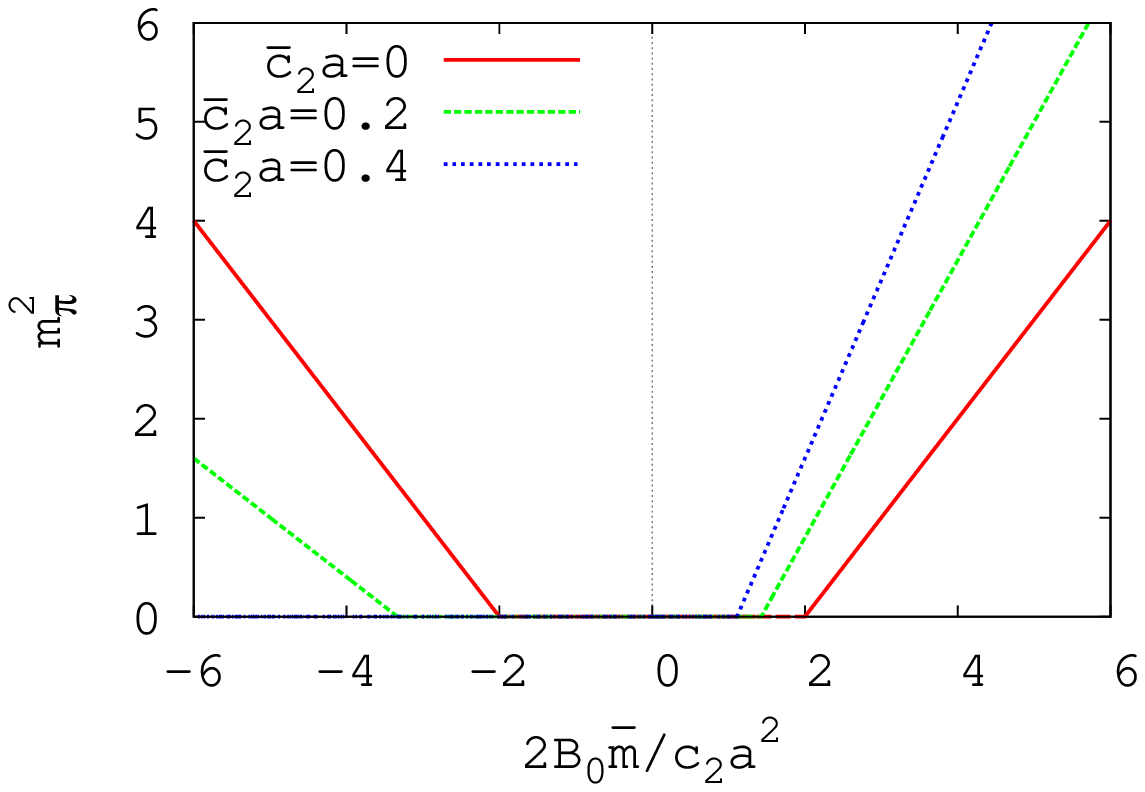}\\
(c) Charged pion mass  for $c_2 > 0$
\end{center}
\end{minipage}
\begin{minipage}{0.5\textwidth}
\begin{center}
\includegraphics[width=0.8\textwidth]{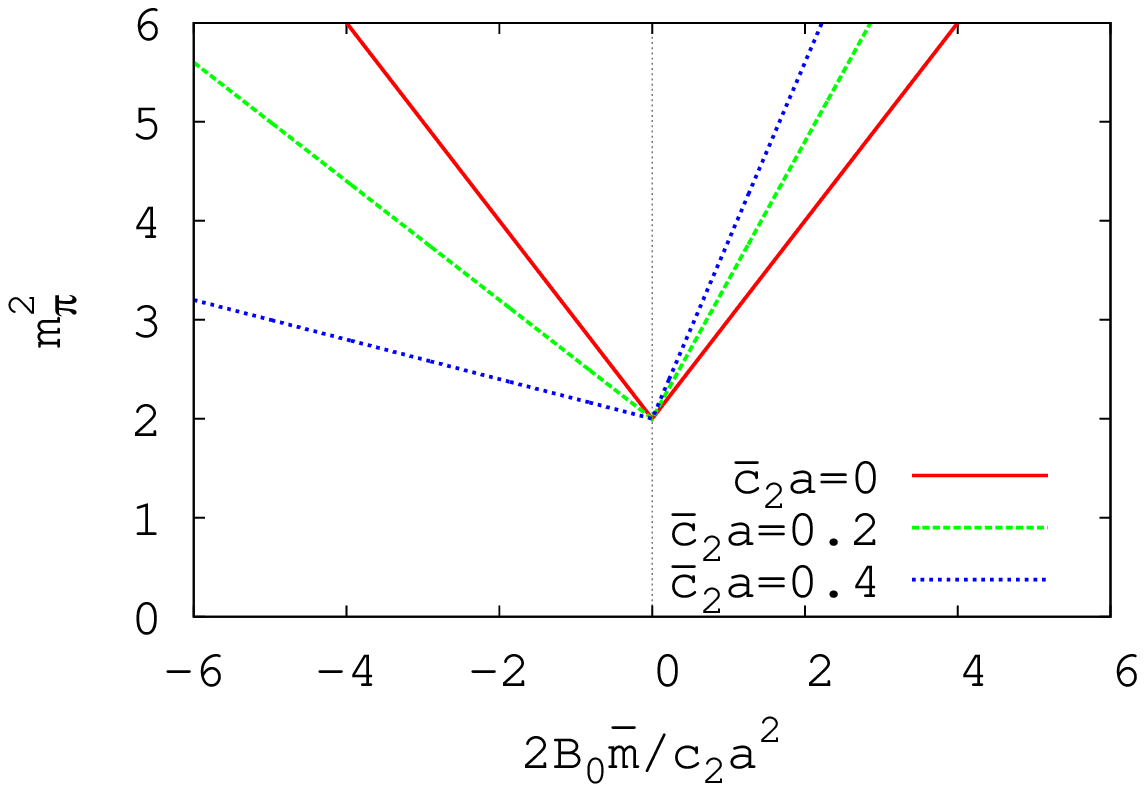}\\
(d) Charged pion mass  for $c_2 < 0$
\end{center}
\end{minipage}
\begin{minipage}{0.5\textwidth}
\begin{center}
\includegraphics[width=0.8\textwidth]{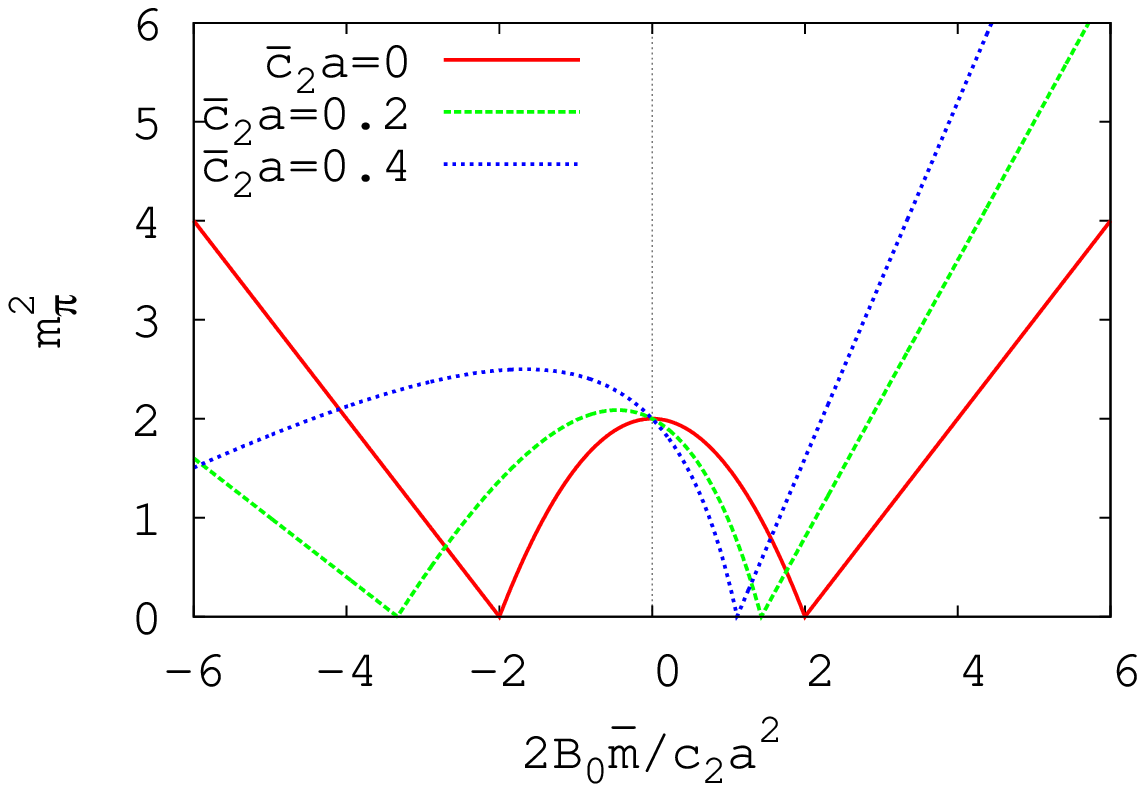}\\
(e) Neutral Pion mass for $c_2 > 0$
\end{center}
\end{minipage}
\begin{minipage}{0.5\textwidth}
\begin{center}
\includegraphics[width=0.8\textwidth]{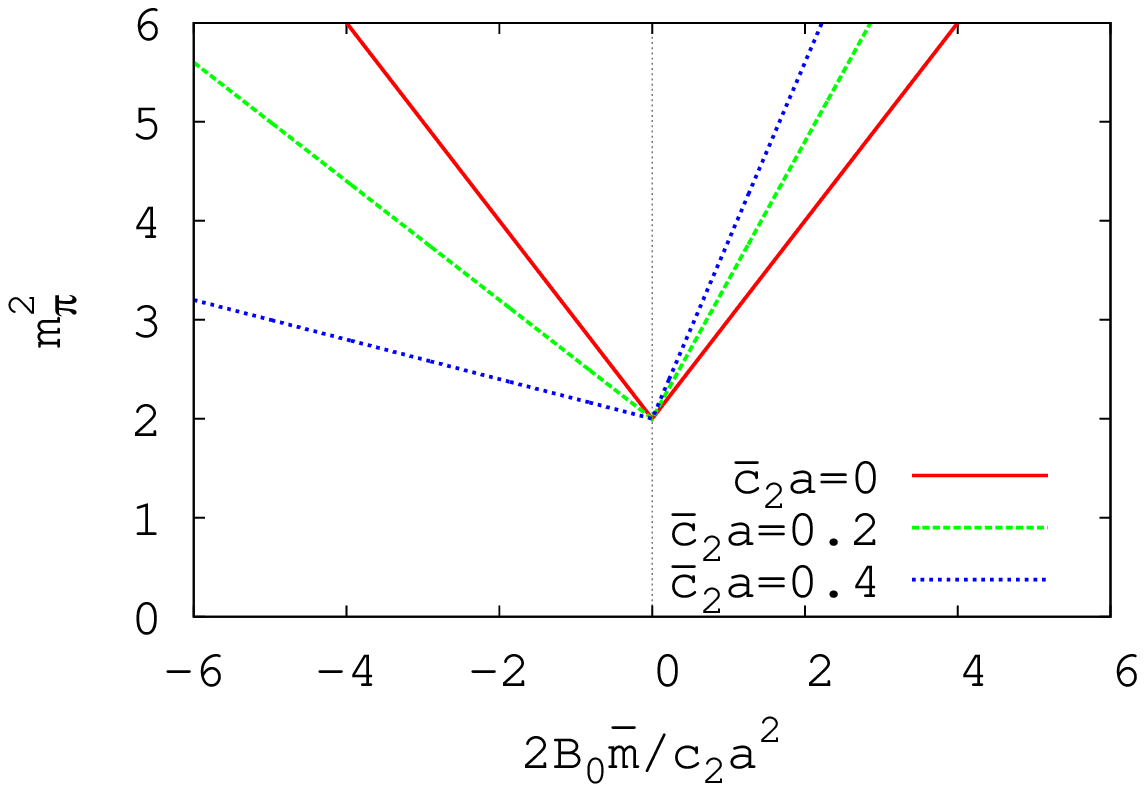}\\
(f) Neutral Pion mass for $c_2 < 0$
\end{center}
\end{minipage}
\caption{$\tilde c_2 a$ dependence at $2B_0\m = 0$}
\label{fig:tc2_dep_mu0}
\end{figure}

These qualitative features remain true even
for the case that $\tilde c_2 \neq 0$.
In figure~\ref{fig:tc2_dep_mu0}(Left) we show results for $c_2 > 0$ at $\mu =0$.
In this case the 2nd order transition points of
the parity-flavor breaking phase move toward more negative values
as the $\tilde c_2 a$ increases. Pion masses are degenerate
and their slopes increase (decrease) in the right(left)
of the phase transition points in the symmetric phase,
while charge pions become massless NG bosons
and the neutral pion is massive in the broken phase.
On the other hand, in the case that $c_2 < 0$ and $\mu =0$,
the first order phase transition line does not move at all
and the pion mass is still degenerate,
as shown in figure~\ref{fig:tc2_dep_mu0}(Right).
The slope of both pion masses increase (decrease)
in the positive (negative) $\tilde m$ region.

\subsection{Vertices}
\label{section:vertices}

Since the twisted mass term $\m$ explicitly break the parity symmetry,
terms including odd powers of pion fields,
$\lag_{\LO + \SLO}^{(2n + 1)}$ with $ n=1,2,\cdots$, exist
in addition to those with even powers, $\lag_{\LO + \SLO}^{(2n)}$.
For $\pi^3$ terms, one finds
\begin{align}
\lag_{\LO + \SLO}^{(3)} &= - \frac{c_0a \sin\phi_0}{2f_0}
\pi_3 (\p_\m\pi_a)^2\notag\\
&+ \frac{1}{6f_0}[2B_0\m'+ 8c_2a^2 \cos\phi_0 \sin\phi_0
- 4\tilde c_2a (2B_0m' \sin\phi_0 - 2B_0\m'\cos\phi_0)]\pi^2 \pi_3,
\label{eq:pi3Lagrangian}
\end{align}
while for the $O(\pi^4)$ Lagrangian we have
\begin{align}
\lag_{\LO + \SLO}^{(4)} &= \frac{1}{6f_0^2}\left[
(\pi_a \p_\m \pi_a)^2 - \left(1 + \frac{3}{2}c_0a\cos\phi_0\right)
\pi^2(\p_\m\pi_a)^2 \right]\notag\\
&-\frac{1}{24f_0^2}[2B_0m' - 8c_2a^2\cos^2\phi_0 + 8\tilde c_2a(2B_0m')\cos\phi_0
](\pi^2)^2\notag\\
&-\frac{1}{24f_0^2}[8c_2a^2\sin^2\phi_0 + 8\tilde c_2a(2B_0\m')\sin\phi_0
]\pi^2 (\pi_3)^2.
\label{eq:pi4Lagrangian}
\end{align}

\subsection{Axial current and ``decay constants'' in twist base}

The quark bilinear operators are defined through the derivatives of
the Lagrangian with respect to source terms as
\begin{align}
S^0 &= \bpsi\psi = \frac{\d}{\d s}\lag,&
P^a &= \bpsi \gamma_5 \frac{\tau^a}{2} \psi
= -\frac{i}{2} \frac{\d}{\d p^a}\lag,\\
V^a_\m &= \bpsi \gamma_\mu \frac{\tau^a}{2} \psi
= \frac{i}{2}\left(\frac{\d}{\d r_\m^a} + \frac{\d}{\d l_\m^a}\right) \lag,&
A^a_\m &= \bpsi \gamma_\mu \gamma_5 \frac{\tau^a}{2} \psi
= \frac{i}{2}\left(\frac{\d}{\d r_\m^a} - \frac{\d}{\d l_\m^a}\right) \lag.
\label{eq:def_ChPT_axial}
\end{align}
Applying these derivatives to the tree-level Lagrangian
(\ref{eq:tree-level_tmWChPT_action}), we obtain the axial current as
\begin{align}
A^a_{\m} &= - \vev{\t^a(\S\p_\m\S^\dag - \S^\dag\p_\m\S)}\\
&= \begin{cases}
if_0\p_\m\pi_a\cos\phi_0 & (a = 1,2)\\
if_0\p_\m\pi_a & (a = 3)
\end{cases},
\end{align}
where in second line, we expand the $\S$ in terms of pion fields $\pi_a$
up to the $O(\pi)$ order.
Since the pseudo scalar ``decay constant'' $f_{PS}$ in twist base is
defined from the expectation value of $A^a_\m$
between the vacuum and one NG boson state as
\begin{equation}
\vev{0| A^a_\m(x) | \pi_b(p)} = f^a_{PS}p_\m e^{-ipx}\d_{ab},
\end{equation}
we obtain
\begin{equation}
f^a_{PS} = \begin{cases}
f_0\cos\phi_0 & (a = 1,2)\\
f_0& (a = 3)
\end{cases}
\end{equation}
at this order.
Note that this decay constants is in twist base, and we need the vector
current for calculate it in physical base.

The $\pi^3$ terms in the axial current $A_\m^a$,
needed at the NLO calculation, is given by
\begin{align}
(A_\mu^{1,2})^{(3)}_{\LO + \SLO} &= -if\left[
\frac{2}{3f_0^2}\pi^2\p_\m\pi_a - \frac{1}{3f_0^2}\pi_a\p_\m\pi^2
+ \frac{c_0a\cos\phi_0}{2f_0^2}\pi^2\p_\m\pi_a
\right]\cos\phi_0\notag\\
&\quad -if\frac{c_0a\sin^2\phi_0}{f_0^2}\pi_3(\pi_a\p_\m\pi_3 - \pi_3\p_\m\pi_a),
\label{eq:A12_pi3}\\
(A_\mu^3)^{(3)}_{\LO + \SLO} &= -if\left[
\frac{2}{3f_0^2}\pi^2\p_\m\pi_a - \frac{1}{3f_0^2}\pi_a\p_\m\pi^2
+ \frac{c_0a\cos\phi_0}{2f_0^2}\pi^2\p_\m\pi_a
\right].\label{eq:A3_pi3}
\end{align}

\section{"next leading order" analysis}
\label{sec:next_leading_order}

\subsection{NLO and NSLO Lagrangian}

The NLO and NSLO Lagrangians are given by
\begin{equation}
\lag_\NLO = \lag_{p^2 \bm{m}, \bm{m}^2},\quad
\lag_\NSLO =  \lag_{ap^2\bm{m},a\bm{m}^2} + \lag_{a^2p^2, a^2\bm{m}}.
\end{equation}
We here do not include the $O(p^4, ap^4)$ terms,
since they do not contribute  to pion masses and decay constants at this order.
The $O(p^2\bm{m}, \bm{m}^2)$ terms are the ordinal NLO terms
in continuum ChPT\cite{Gasser:1983ky,Gasser:1983kx,Gasser:1983yg, Gasser:1984gg} and are given by
\begin{equation}
\lag_{p^2\bm{m},\bm{m}^2} = L_{45} \vev{\S \tchi^\dag + \tchi\S^\dag}\vev{D_\m\S D_\m\S^\dag}
- L_{68}\vev{\S \tchi^\dag + \tchi\S^\dag}^2 .
\end{equation}
In $O(p^2\bm{m}, \bm{m}^2)$ terms, we follow the notation of
\cite{Sharpe:2004ny}. The relation of the low energy constants
in the notations between \cite{Gasser:1983yg} $l$ and
\cite{Sharpe:2004ny} $L$ is given by
$L_{45} = l_4/8, L_{68} = (l_3 + l_4)/16$.

The $O(ap^2, a \bm{m}^2)$ and $O(a^2p^2, a^2 \bm{m})$ terms
describe the lattice artifacts and they are constructed
as usual using the spurion analysis as
\begin{align}
\lag_{ap^2\bm{m},a\bm{m}^2} &=
\vev{D_\m\S D_\m\S^\dag}\left[
V_1 \vev{\S\hata^\dag + \hata\S^\dag}\vev{\S \tchi^\dag + \tchi\S^\dag}
+ V_2\vev{\hata\tchi^\dag + \tchi\hata^\dag}
\right]\notag\\
&+ V_3\vev{D_\m\S\hata^\dag + \hata D_\m\S^\dag}
\vev{D_\m\S\tchi^\dag + \tchi D_\m\S^\dag}\notag\\
&+ V_4 \vev{\S\hata^\dag + \hata\S^\dag}\vev{\S \tchi^\dag + \tchi\S^\dag}^2
+ V_5 \vev{\tchi \tchi^\dag}\vev{\S\hata^\dag + \hata\S^\dag}\notag\\
&+ V_6\vev{\hata\tchi^\dag + \tchi\hata^\dag}
\vev{\S \tchi^\dag + \tchi\S^\dag}\notag\\
&+ \tV_{23} \vev{D_\m\S\hata^\dag D_\m\S\tchi^\dag
+ D_\m\S^\dag \tchi D_\m\S^\dag\hata}
+ \tV_{3a} \vev{D^2\S\hata^\dag + \hata D^2\S^\dag}
\vev{\S \tchi^\dag + \tchi\S^\dag}\notag\\
&+ \tV_{3m} \vev{D^2\S \tchi^\dag + \tchi D^2\S^\dag}
\vev{\S\hata^\dag + \hata \S^\dag},
\end{align}
\begin{align}
\lag_{a^2p^2,a^2\bm{m}} &=
\vev{D_\m\S D_\m\S^\dag}\left[
X_1 \vev{\S\hata^\dag + \hata\S^\dag}^2 + X_2\vev{\hata\hata^\dag}\right]
+ X_3\vev{D_\m\S\hata^\dag + \hata D_\m\S^\dag}^2\notag\\
&+ X_4 \vev{\S\hata^\dag + \hata\S^\dag}^2\vev{\S \tchi^\dag + \tchi\S^\dag}
+ X_5 \vev{\hata\hata^\dag}\vev{\S\tchi^\dag + \tchi\S^\dag}\notag\\
&+ X_6\vev{\hata\tchi^\dag + \tchi\hata^\dag}
\vev{\S \hata^\dag + \hata\S^\dag}\notag\\
&+ \tX_{23}\vev{D_\m\S\hata^\dag D_\m\S\hata^\dag
+ D_\m\S^\dag \hata D_\m\S^\dag\hata}\notag\\
&+ \tX_{3} \vev{D^2\S\hata^\dag + \hata D^2\S^\dag}
\vev{\S \hata^\dag + \hata\S^\dag},
\end{align}
where the $V_i (i = 1,\dots,6), \tV_i (i = 23, 3a, 3m),
X_i (i =1,\dots, 6),$ and $\tX_i, (i= 23, 3)$
are additional low energy constants.

Here we introduce the nineteen parameters in NLO Lagrangian
but our aim is not to determine these parameters.
We instead investigate quark and lattice spacing dependences
of physical observables using these parameters.
In the physical observable, these parameters always appear
as the some linear combinations,
so that the number of independent fit parameters are much smaller.

Expanding $\S$ in terms of component fields $\pi$,
$O(\pi)$ and $O(\pi^2)$ terms of NLO+NSLO Lagrangian
$\lag_{\NLO+\NSLO}^{(1)}$ and $ \lag_{\NLO+\NSLO}^{(2)}$ are given by
\begin{align}
\lag_{\NLO+\NSLO}^{(1)}
 &= C^1_\NLO \pi_3,\label{eq:NOL_for_vacuum}\\
\lag_{\NLO+\NSLO}^{(2)}
 &= \frac{1}{2}\left(C^{2p,a}_\NLO (\p_\m\pi_a)^2 + C_\NLO^{2,a}\pi_a^2\right).
\label{eq:NLO_for_mass}
\end{align}
The coefficients $C_\NLO$'s are given by
\begin{align}
C^1_\NLO &= \frac{8}{f_0}\Bigl[
- (2B_0m')(2B_0\m')4L_{68} - a(2B_0m')^2(8V_4' + V_5')\sin\phi_0\notag\\
&\qquad + a(2B_0m')(2B_0\m')(16V_4'+2V_6')\cos\phi_0
- a(2B_0\m')^2(V_5' + 2V_6')\sin\phi_0\notag\\
&\qquad - a^2(2B_0m')(8X'_4 + X'_6)\sin 2\phi_0\notag\\
&\qquad + a^2(2B_0\m')\left\{(4X'_4 + X'_5 + X'_6) + (4X'_4 - X'_6)\cos
 2\phi_0\right\}\Bigr],
\end{align}
\begin{align}
C_\NLO^{2p,a} &= \frac{16}{f_0^2}\Bigl[
2B_0m' L_{45} + a\left\{2B_0m'(4V'_1 +V'_2 - \tV'_{23}/2)\cos\phi_0
- 2B_0\m'(V'_2 - \tV'_{23}/2) \sin\phi_0\right\}\notag\\
&\qquad + a^2\left\{(2X'_1 + X'_2/2 - \tX'_{23}/2)
+ 2X'_1\cos 2\phi_0\right\}\notag\\
&\qquad - \Bigl\{a(2B_0\m')(2V'_3 + \tV'_{23}
- 2\tV'_{3a} - 2\tV'_{3m})\sin\phi_0 \notag\\
&\qquad\qquad - a^2(X'_3 + \tX'_{23}/2 - \tX'_3) (1 - \cos 2\phi_0)\Bigr\}\d_{a3}
\Bigr],
\end{align}
\begin{align}
C_\NLO^{2,a} &= -\frac{16}{f^2_0}\Bigl[
- (2B_0m')^2 2L_{68} + a \bigl\{ (2B_0m')^2(12V'_4 + V'_5/2 + V'_6)
\cos\phi_0\notag\\
&\quad \qquad \qquad - (2B_0m')(2B_0\m')V'_6\sin\phi_0
+ (2B_0\m')V'_5/2 \cos\phi_0\bigr\}\notag\\
&\qquad + a^2(2B_0m')\left\{(6X'_4 + X'_5/2 + X'_6/2)
+ (6X'_4 + X'_6/2)\cos 2\phi_0
- a^2 (2B_0\m')X'_6/2 \sin 2\phi_0\right\}\notag\\
&\qquad + \bigl\{(2B_0\m')^2 2L_{68} - a 8V'_4 \{ (2B_0\m')^2 \cos \phi_0
+ 2(2B_0m')(2B_0\m')\sin\phi_0\}\notag\\
&\qquad \qquad - 4X'_4 a^2 \left\{(2B_0m')(1 - \cos 2\phi) - 2(2B_0\m')
\sin 2\phi_0\right\}
\bigr\}\d_{3a}\Bigr],
\end{align}
where we use the normalized coefficients $V'_i = 2W_0 V_i$,
$\tV'_i = 2W_0 \tV_i$, $X'_i = (2W_0)^2 X_i$ and $\tX'_i = (2W_0)^2 \tX_i$.
Note again that, since the twisted mass term explicitly breaks
the parity symmetry, odd power terms appear.
In particular, the $O(\pi)$ terms, together with 1-loop contributions,
must be canceled by the redefinition of the vacuum angle $\phi_0$ as
\begin{equation}
\phi_R = \phi_0 + \D\phi,
\end{equation}
where $\D\phi$ is the NLO contribution.

\subsection{The vacuum expectation value and pion masses}

Due to the presence of three-point vertices (\ref{eq:pi3Lagrangian}),
the tadpole diagrams $\lag_\Loop^{(1)}$ contribute to
vacuum expectation value at 1-loop,
\begin{align}
 \lag_\Loop^{(1)} &= \frac{1}{f_0}(A_\pm I_\pm + A_0 I_0)\pi_3,
 \label{eq:1loopfortadpole1}\\
 A_\pm &= c_2 a^2 \sin 2\phi_0 + (c_0 - \tilde c_2) a (2B_0m') \sin\phi_0,
 \label{eq:1loopfortadpole2}\\
 A_0 &= \frac{3}{2}c_2 a^2 \sin 2\phi_0 + \frac{1}{2}(c_0 - 3\tilde c_2)
 a (2B_0m') \sin\phi_0,
 \label{eq:1loopfortadpole3}
\end{align}
where $I_a$ is given by
\begin{equation}
 I_a = \int \frac{d^4p}{(2\pi)^4} \frac{1}{p^2 + (m_\pi^a)^2}.
\end{equation}
See the  detail of the calculation in Appendix \ref{sec:app_detail_calc}.
This contribution diverges and therefore must be
renormalized by $O(\pi)$ terms in $\lag_{\NLO+\NSLO}^{(1)}$,
together with the redefinition of the vacuum angle $\phi_R$.
Explicitly the renormalization condition becomes
\begin{equation}
\left.\frac{\p}{\p \phi}\lag_{\LO + \SLO}^{(1)}\right|_{\phi = \phi_0} \D\phi
+ \lag_\Loop^{(1)} + \lag_{\NLO+\NSLO}^{(1)} = 0,
\end{equation}
which leads to
the renormalized vacuum angle $\phi_R=\phi_0+\Delta\phi$ as
\begin{equation}
\phi_R = \phi_0 - \frac{1}{(m_\pi^0)^2} \left[
A_\pm L_\pm + A_0 L_0 - \frac{8}{f_0^2}\left\{
a(2B_0m')^2C_1 + a^2(2B_0m')C_2
\right\}
\right].
\end{equation}
The first term is the tree-level contribution for vacuum angle,
while the second and third terms are chiral logarithm contributions from
the charged and neutral pion, respectively.
The last two terms are  the polynomial contributions from the NLO Lagrangian.
The chiral logarithm for the pion loop is defined by
\begin{equation}
L_a = \frac{(m_\pi^a)^2}{16\pi^2f_0^2}\log\left(\frac{m_\pi^a}{\m_\ChPT}\right)^2,
\label{eq:chiral_logarithm}
\end{equation}
and the coefficients of the chiral logarithm, $A_\pi^a$, are given
in eqs.(\ref{eq:1loopfortadpole2}) and (\ref{eq:1loopfortadpole3}).
The coefficients of the NLO polynomial term, $C_1$ and $C_2$, are
some combinations of the renormalized NLO low energy constants.

In the similar way, we obtain the result for the pion masses as
\begin{align}
(m_\pi^a)^2_\NLO &= \left.(m_\pi^a)^2_\LO\right|_{\phi_0 \to \phi_0 + \D\phi}\notag\\
& + \sum_{b = \pm, 0}\left( m_\pi^2 B^a_b + a^2 Q^a_b \right)L_b\notag \\
&-\frac{16}{f_0^2} \left\{E_1^a (2B_0m')^2 + E_2^a a(2B_0m')^2+ E_3^a a^2(2B_0m')\right\}.
\label{eq:NLO_mass_formulation}
\end{align}
This is one of the main results of this paper.
The first term is the tree-level form for pion mass
but with the vacuum angle obtained at NLO.
The second line represents the chiral logarithm while
the polynomial terms for the NLO Lagrangian are given in the last line.
Each coefficient of the chiral logarithms is given by
\begin{align}
B_\pm^\pm &= a(2c_0 - 4 \tilde c_2)\cos\phi_0,&
B_0^\pm &= \frac{1}{2} + a(c_0 - \tilde c_2) \cos\phi_0,
\label{eq:constA}\\
B_\pm^0 &= 1 + a(2c_0 - 2\tilde c_2)\cos\phi_0,&
B_0^0 &= - \frac{1}{2} + a(c_0 - 3\tilde c_2)\cos\phi_0,
\label{eq:constB}\\
Q_\pm^\pm &= 4c_2 \cos^2\phi_0,& Q_0^\pm &= c_2 \cos^2\phi_0,
\label{eq:constC}\\
Q_\pm^0 &= 2 c_2 \cos^2\phi_0,& Q_0^0 &= c_2 (3 \cos^2\phi_0 - 4 \sin^2\phi_0).
\label{eq:constD}
\end{align}
and coefficients of polynomial terms $E_i (i = 1,2,3)$
are combinations of the renormalized NLO low energy constants.
Our results have not only the same pion mass contributions
for the chiral logarithms in \cite{Bar:2010jk},
but also the additional $O(a\bm{m})$ terms from SLO.
Note that $O(1)$ terms in eqs.(\ref{eq:constA}) and (\ref{eq:constB}) of course agree with
continuum results \cite{Gasser:1983ky,Gasser:1983kx}.

\subsection{''Decay constants''}

Let us consider the NLO axial currents.
Using eq. (\ref{eq:def_ChPT_axial}),
we obtain the "NLO"(=NLO+NSLO) terms of the axial currents for $a = 1,2$,
\begin{align}
(A_\m^{a})_\NLO &= \frac{16i}{f_0}\p_\m \pi_a \Bigl[\bigl\{
(2B_0m')L_{45} + a(2B_0m')(4V'_1 + V'_2 - \tV'_{23}/2
- 2 \tV'_{3m})\cos\phi_0\notag\\
& \qquad - a(2B_0\m')(V'_2 - \tV'_{23}/2 - 2\tV'_{3m})\sin\phi_0\notag\\
& \qquad + a^2(2X'_1 + X'_2/2 - \tX'_{23}/2 - 2 \tX'_3 + 2 X'_1 \cos 2\phi_0)
\bigr\} \cos\phi_0 \notag\\
& \qquad-a(2B_0m')2\tV'_{3a}\Bigr],
\end{align}
and for $a = 3$,
\begin{align}
(A_\m^{3})_\NLO &= \frac{16i}{f_0}\p_\m \pi_3 \Bigl[\bigl\{ (2B_0m')L_{45}
+ a(2B_0m')(4V'_1 + V'_2 - \tV'_{23}/2 -2 \tV'_{3a} - 2 \tV'_{3m})
\cos\phi_0\notag\\
& \qquad - a(2B_0\m')(V'_2 + 2V'_3 + \tV'_{23}/2)\sin\phi_0\notag\\
& \qquad + a^2\bigl\{ 2X'_1 + X'_2/2 + X'_3- \tX'_3
+ (2 X'_1 - X'_3 - \tX'_{23}/2 - \tX'_3)\cos 2\phi_0\bigr\}\Bigr].
\end{align}

From these results, we calculate the ``decay constants'' for
the pseudo-scalar meson at "NLO", which are
given by
\begin{align}
f_{PS,\NLO}^a &= \left.f_{PS,\LO}^a\right|_{\phi = \phi_0 + \D\phi}
+ f^a_{PS, \LO}\Bigl[
\sum_{b = \pm,0} F_b^a L_b\notag\\
& \qquad+ \frac{16}{f_0^2} \left\{
(2B_0m') H_1^a + a(2B_0m') H_2^a + a^2 H_3^a\right\}\Bigr]\notag\\
& \qquad + \D f_\pm (1 - \d_{3a}).
\label{eq:decay_const}
\end{align}
This is also one of the main results of this paper.
The first term is the tree-level contribution, which includes
the vacuum renormalization effects at "NLO".
The second term represents the chiral logarithms and
the polynomial terms for the NLO Lagrangian are in the second line.
In addition, the ``decay constant'' for the charged pseudo scalar has
the additive renormalization term in the last term.
Coefficients of the chiral logarithms are given by
\begin{align}
F_\pm^\pm &= - \frac{1}{2}\left(1 + c_0a\cos\phi_0\right),&
F_0^\pm &= - \frac{1}{2}\left(1 + \frac{5}{2}c_0a\cos\phi_0\right),
\label{eq:const2}\\
F_\pm^0 &= - \left(1 + \frac{1}{2} c_0a\cos\phi_0\right),&
F_0^0 &= - \frac{1}{4}c_0 a\cos\phi_0,
\label{eq:const3}
\end{align}
and the formula for the coefficients of the renormalization terms
$H_i (i = 1,2,3)$ and $\Delta f_\pm$ are given
in Appendix \ref{sec:app_detail_calc}.
Note again that $O(1)$ terms in eqs.(\ref{eq:const2}) and
(\ref{eq:const3}) agree with continuum results
\cite{Gasser:1983ky,Gasser:1983kx}.

\subsection{Consistency}

Divergences generated by 1-loop contributions must be removed by the "NLO" terms.
For this purpose we define the renormalized NLO and NSLO low energy constants as
\begin{align}
L_i &= L^r_i(\m_\ChPT) + \frac{l_i}{32\pi^2}R, \\
V'_i &= V^r_i(\m_\ChPT) + \frac{v_i}{32\pi^2}R, \\
X'_i &= X^r_i(\m_\ChPT) + \frac{x_i}{32\pi^2}R,
\end{align}
where $l_i$, $v_i$, $x_i$ are renormalize parameters,
the argument $\m_\ChPT$ represents the renormalized scale
and $R=O(1/\epsilon)$ is  introduce to cancel  1-loop divergences in the dimensional regularization.
We give the detail formulation in appendix.~\ref{sec:app_detail_calc}

The renormalization conditions for the vacuum expectation value leads to
\begin{align}
8(8v_4 + v_5 + 4 l_{68}\tilde c_2) = 3c_0 - 5 \tilde c_2,
\label{eq:rcondition_v1}\\
8(8x_4 + x_6 - 4 l_{68} c_2) = 5  c_2.
\label{eq:rcondition_v2}
\end{align}

In order to cancel the divergences in 1-loop diagrams for pion masses,
coefficients should satisfy the following conditions.
\begin{align}
16(l_{45} - 2l_{68}) &= 1,\label{eq:rcondition_mass1}\\
16(12v_4 + v_5/2 + v_6 + 4v_1 + v_2 - \tilde v_{23}/2 + 2\tilde c_2 l_{45})
&= -6\tilde c_2 + 6c_0,\\
16(6x_4 + x_5/2 + x_6/2 + 2x_1 + x_2/2 - \tilde x_{23}/2 - c_2l_{45}) &= 4c_2,\\
16(6x_4 + x_6/2 + 2x_1 - c_2l_{45}) &= 2c_2,\\
16(4x_4 -  x_3 - \tilde x_{23}/2 + \tilde x_3 - c_2l_{45}) &= - 6c_2.
\label{eq:rcondition_mass2}
\end{align}

Similarly from the decay constant, we have
\begin{align}
4l_{45} &= 1,\\
16\tilde v_{3a} &= c_0,\\
8(2v_1 + v_2/2 - \tilde v_{23}/4 - 2\tilde v_{3m}) &= 2\tilde c_2 + 7c_0/4,\\
16x_1 &= - 3c_2,\\
16(x_2/4 - \tilde x_{23}/4 - 2 \tilde x_3) &= 2c_2,\\
16(x_3/2 + \tilde x_{23}/4 + 3 \tilde x_3/2) &= - c_2.
\end{align}

A solutions to these conditions can be obtained as
\begin{align}
 l_{45} &= \frac{1}{4},& l_{68} &= \frac{3}{32},\\
 x_1  &= - \frac{3}{16}c_2 ,& x_2 - \tilde x_{23} &= - c_2,&
 x_3  + \frac{\tilde x_{23}}{2}&= \frac{7}{16} c_2,\\
 x_4  &= \frac{1}{8}c_2,& x_5 &= \frac{5}{4}c_2,& x_6 &= 0,\\
 \tilde x_3 &= - \frac{3}{16} c_2,
\end{align}
\begin{align}
 16v_1 + 4v_2 -2\tilde v_{23} - 16\tilde v_{3m}
 &= \frac{7}{4}c_0 + 2\tilde c_2, \\
 64v_4 + 8v_5 &= 3c_0 - 8\tilde c_2,\\
 64v_4 + 8v_6 + 32 \tilde v_{3m} &= -2c_0 - 7 \tilde c_2,\\
 \tilde v_{3a} &= \frac{c_0}{16}.
\end{align}

Note that results of $l_{45}$ and $l_{68}$ agree with the continuum ChPT
\cite{Gasser:1983ky,Gasser:1983kx}.

%
%
\section{maximal twist}
\label{sec:maximal_twist}

\begin{figure}[htbp]
\begin{center}
\includegraphics[width=.5\textwidth]{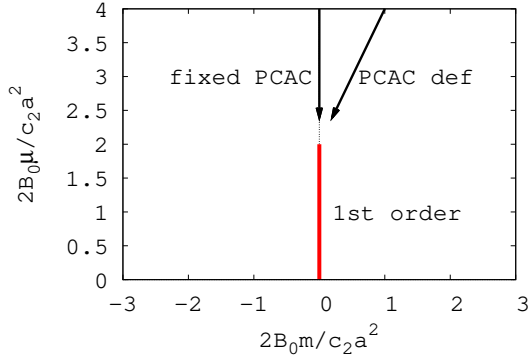}
\caption{The chiral limit for $c_2 < 0$,
where the first order phase transition line exists.
The untwisted mass $m$ corresponding to the maximal twist depends on
twisted mass $\m$ in the PCAC definition,
while $m$ is constant in the fixed PCAC definition.}
\label{fig:climit_1st}
\end{center}
\end{figure}

One of the advantage of the tmlQCD  is
the automatic $O(a)$ improvement at maximal twist
\cite{Frezzotti:2001ea, Aoki:2006nv}.
The ETMC employs the PCAC quark mass to determine the maximal twist
in their simulation,
\begin{equation}
m_\PCAC (\mu) = \frac{\sum_{\bm{x}}\vev{\p_4A_4^a(\bm{x},t)P^a(0)}}
{2\sum_{\bm{x}}\vev{ P^a(\bm{x},t) P^a(0)}} = 0 \qquad (a = 1,2).
\end{equation}
The above condition must be satisfied by tuning the untwisted mass $m$
at each twisted mass $\mu$\cite{Aoki:2006nv}.
The untwisted mass which realizes the maximal twist condition is called
the critical untwisted quark mass.
This definition for the maximal twist is called the PCAC definition.
In order to reduce the numerical cost, however,
$m$ for the maximal twist  is defined at $\mu_{\rm min}$,
the minimal value of the twisted mass employed in the simulations,
such that $m_{PCAC}(\mu_{\rm min})=0$.
This is called as the fixed PCAC definition.
In figure \ref{fig:climit_1st},
both definitions are schematically drawn for $c_2 < 0$.
In this section, we investigate the difference
between two maximal twist conditions, using our WChPT analysis.

\subsection{PCAC quark mass}

The PCAC quark mass can be written in the WChPT as
\begin{equation}
m_\PCAC = \frac{f_{PS} m_\pi^2}{2Z(0)},
\end{equation}
where $f_{PS}, m_\pi$ denote the decay constant and mass for the charged pion,
and we define $Z(p)$ as
\begin{equation}
\vev{0| P^a(x) | \pi(p)} = i Z(p) e^{-ipx}.
\end{equation}
At "NLO",  we have
\begin{align}
f_{PS,\NLO} &= \left. f_{PS,\LO} \right|_{\phi_0 \to \phi_0+\D\phi}
(1 + \D_f) + \D f_\pm,\\
m_{\pi,\NLO}^2 &= \left. m_{\pi,\LO}^2 \right|_{\phi_0 \to \phi_0+\D\phi}
(1 + \D_m),\\
Z_{\NLO}(p) &= \left. Z_{\LO}(p) \right|_{\phi_0 \to \phi_0+\D\phi} (1 + \D_Z),
\end{align}
where the $\D_i (i = f,m,Z)$ denote the NLO contribution corresponding to
$f_{PS}, m_\pi$ and $Z(p)$.
The PCAC quark mass at the NLO is given by
\begin{align}
m_{\PCAC,\NLO} &= \left.m_{\PCAC,\LO}\right|_{\phi_0 \to \phi_0+ \D\phi}\left(
1 + \D_f + \D_m - \D_Z\right),\notag\\
&+ \left.\frac{m^2_{\pi,\LO}}{2Z_\LO(0)}\right|_{\phi_0 \to \phi_0+\D\phi}
\D f_\pm
\label{eq:NLO_pcac}
\end{align}
where
\begin{equation}
\left.m_{\PCAC,\LO}\right|_{\phi_0 \to \phi_0+ \D\phi}
= \frac{m_{\pi,\LO}^2}{2B_0(1 + \tilde c_2 \cos(\phi_0+\D\phi))} \cos(\phi_0+\D\phi).
\end{equation}

\subsection{PCAC definition}

The maximal twist condition in the PCAC definition leads to
\begin{equation}
\cos \phi_0 = O(am') = O(a\mu).
\label{eq:PCAC_def_mt}
\end{equation}
This condition simplify the gap equation (\ref{eq:WChPT_gap_eq}),
and the critical untwisted quark mass  as a function of
the twisted mass $\m$ is given by
\begin{equation}
2B_0\tilde m = - \tilde c_2 a 2B_0\m + O(a\mu^2).
\label{eq:critical_mass}
\end{equation}
Eq. (\ref{eq:critical_mass}) shows that
the critical untwisted quark mass $\tilde m$ depends linearly
on the twisted quark mass $\m$.
Using eqs. (\ref{eq:PCAC_def_mt}) and (\ref{eq:critical_mass}),
the charged pion mass at the NLO is given by
\begin{equation}
(m_\pi^\pm)^2_\NLO = (m_\pi^\pm)^2_\LO\left[
1 + \frac{1}{2}L_0
- \frac{16}{f^2_0}\left( (m_\pi^\pm)^2_\LO L_c^r + a^2 X_c^r\right)
\right],
\label{eq:charged_mass_at_maximal_twist}
\end{equation}
where, the NLO polynomial coefficients $L_c^r$ and $X^r_c$ are given by
\begin{align}
L_c^r &= L_{45}^r - 2L_{68}^r,\label{eq:maximaltwist_lec_1}\\
X^r_c &= (X_5^r /2 + X_2^r/2 - \tX_{23}^r/2),\label{eq:maximaltwist_lec_2}
\end{align}
the chiral logarithm $L_0$ is defined by eq. (\ref{eq:chiral_logarithm}),
and the LO pion masses are simplified as
\begin{align}
(m_\pi^\pm)^2 &= 2B_0\m,\\
(m_\pi^0)^2 &= 2B_0\m + a^2(2c_2 + 2\tilde c_2^22B_0\m).
\end{align}
Eq. (\ref{eq:charged_mass_at_maximal_twist}) shows that
the charged pion mass is $O(a)$ improved:
the lattice spacing corrections start at $O(a^2)$.
It is important to note that the charged pion mass
at the NLO contains the chiral logarithm only from the neutral pion loops
but not from the charged pion loops at the maximal twist,
as in the continuum ChPT.

Note that the maximal twist by the PCAC quark mass and the charged axial
currents give the same constraint for physical observable in this WChPT analysis.

\subsection{Fixed PCAC definition}

For the fixed PCAC definition, the maximal twist condition
and the critical untwisted quark mass are given by
\begin{align}
\cos \phi_0 &= O(a\mu_{\rm min}) 
\label{eq:fixed_PCAC_mt}\\
2B_0\tilde m &= - \tilde c_2 a 2B_0\m_{\rm min} + O(a\mu_{\rm min}^2), 
\label{eq:critical_mass_fixed_PCAC}
\end{align}
where $\m_{\rm min}$ denotes the minimal value of the twisted mass,
used for the fixed PCAC definition.
Eq. (\ref{eq:critical_mass_fixed_PCAC}) shows that
the critical untwisted quark mass is constant
and does not depend on the twisted mass $\mu$.
Using eqs. (\ref{eq:fixed_PCAC_mt}) and (\ref{eq:critical_mass_fixed_PCAC}),
the charged pion mass at the NLO is given by
\begin{equation}
(m_\pi^\pm)^2_\NLO = \left.(m_\pi^\pm)^2_\LO\right|_{\phi_0 \to \phi_0 + \D\phi}
\left[ 1 + \frac{1}{2}L_0
- \frac{16}{f^2_0}\left( (m_\pi^\pm)^2_\LO L_c^r + a^2 X_c^r\right)
\right],
\label{eq:charged_mass_fixed_PCAC}
\end{equation}
where the coefficients $L_C^r$ and $X_c^r$ are defined
in eqs. (\ref{eq:maximaltwist_lec_1}) and (\ref{eq:maximaltwist_lec_2})
and the pion masses are given by
\begin{align}
\left.(m_\pi^\pm)^2\right|_{\phi_0 \to \phi_0 + \D\phi} &= 2B_0\m,\\
\left.(m_\pi^0)^2\right|_{\phi_0 \to \phi_0 + \D\phi}  &= 2B_0\m
+ a^2(2c_2 + 2\tilde c_2^22B_0\m_{\rm min}).
\end{align}
The charged pion mass (\ref{eq:charged_mass_fixed_PCAC})
for the fixed PCAC definition
has the same functional form as (\ref{eq:charged_mass_at_maximal_twist})
for the PCAC definition.
The fixed critical untwisted quark mass effect, however,  appears
in the $O(a^2)$ terms  for the LO neutral pion mass.

It turns out that the charged pion mass is $O(a)$ improved
for both PCAC and fixed PCAC definitions,
and that the difference between these two definitions appears
as the $O(a^2)$ effects.
In addition, contrary to the case at the non-maximal twist where
the chiral logarithm of the charge pion mass contains effects
from both neutral and  charged pion loops due to the lattice artifact,
it contains only the effect from the neutral pion loop
for both definitions as in the continuum ChPT.

\section{Conclusions}
\label{sec:Conclustion}

In this paper, we construct the Wilson chiral perturbation theory
for the $N_f = 2$ twisted mass lattice QCD at the small quark mass regime
such that $m_q \sim a^2 \Lambda^3$. In order to consider such a regime,
we include $O(a^2)$ and $O(am)$ terms at the tree level
as the sub-leading order Lagrangian,
which induce the non-trivial phase structure
and pion mass splitting at the tree-level.
Using this effective theory, we investigate the pion mass and decay constant
as a function of not only the twisted quark mass but also the lattice cutoff
at the next leading order.
Our main results are given in Eqs. (\ref{eq:NLO_mass_formulation}) and
(\ref{eq:decay_const}).
We also confirm that divergences from 1-loop contributions can be
consistently removed by  the next leading order Lagrangian.

For the comparison of our results with data obtained by numerical simulations,
we derive the twisted quark mass dependence of the charged pion mass
at the maximal twist.
As the definition of the maximal twist, we adopt two different definitions,
the PCAS and the fixed PCAC definition, the latter of which is
actually employed in the simulations.
We have found that the charged pion mass is $O(a)$ improved, so that
lattice spacing corrections start at $O(a^2)$ for both two definitions,
and that the difference between the two definitions appears as $O(a^2)$ effects.
In addition, it should be noted that the chiral logarithm
in the charged pion mass comes from the neutral pion loop only,
as in the continuum ChPT.

\section*{Acknowledgements}
We thank for the important comments form Oliver Bar.
This work is supported in part by the Grant-in-Aid of MEXT(No. 20340047)
and by Grant-in-Aid for Scientific Research on Innovative Areas (No
2004: 20105001,20105003).

\appendix
\section{detail of calculations}
\label{sec:app_detail_calc}

\subsection{NLO terms}
From the spurion analysis,  $O(ap^2, a \bm{m}^2)$ terms are given by
\begin{align}
&O(ap^2\bm{m})&
&\vev{\S\hata^\dag + \hata\S^\dag}\vev{\S \tchi^\dag + \tchi\S^\dag}
\vev{D_\m\S D_\m\S^\dag},
\vev{\hata\tchi^\dag + \tchi\hata^\dag}\vev{D_\m\S D_\m\S^\dag},\notag\\
&&&\vev{D_\m\S\hata^\dag + \hata D_\m\S^\dag}
\vev{D_\m\S\tchi^\dag + \tchi D_\m\S^\dag},
\underbrace{\vev{D_\m\S\hata^\dag D_\m\S\tchi^\dag
+ D_\m\S^\dag \tchi D_\m\S^\dag\hata}}_{S_1},\notag\\
&&&\underbrace{\vev{D^2\S\hata^\dag + \hata D^2\S^\dag}
\vev{\S \tchi^\dag + \tchi\S^\dag}}_{S_2},
\underbrace{\vev{D^2\S \tchi^\dag + \tchi D^2\S^\dag}
\vev{\S\hata^\dag + \hata \S^\dag}}_{S_3},\\
&O(a\bm{m}^2)&
&\vev{\S\hata^\dag + \hata\S^\dag}\vev{\S \tchi^\dag + \tchi\S^\dag}^2,
\vev{\tchi \tchi^\dag}\vev{\S\hata^\dag + \hata\S^\dag},
\vev{\hata\tchi^\dag + \tchi\hata^\dag}\vev{\S \tchi^\dag + \tchi\S^\dag}.
\end{align}
Here $S_1, S_2,$ and $S_3$ terms come from the non-commutativity between
the covariant derivative $D_\m$ and mass term $\chi$.
Therefore they vanish for the degenerated $N_f = 2$
(untwisted) Wilson fermion case.
In the $l_\m = r_\m = 0$ limit, these terms can be expressed by other terms,
\begin{align}
S_1 &\to \frac{1}{2}\left(
\vev{\p_\m\S\hata^\dag + \hata \p_\m\S^\dag}
\vev{\p_\m\S\tchi^\dag + \tchi \p_\m\S^\dag}
- \vev{\hata\tchi^\dag + \tchi\hata^\dag}\vev{\p_\m\S \p_\m\S^\dag}
\right),\\
S_2 &\to - \vev{\p_\m\S\hata^\dag + \hata \p_\m\S^\dag}
\vev{\p_\m\S\tchi^\dag + \tchi \p_\m\S^\dag},\\
S_3 &\to - \vev{\p_\m\S\hata^\dag + \hata \p_\m\S^\dag}
\vev{\p_\m\S\tchi^\dag + \tchi \p_\m\S^\dag}.
\end{align}

In a similar way to $O(a p^2\bm{m}, a\bm{m}^2)$ terms,
$O(a^2p^2, a^2 \bm{m})$ terms are given by
\begin{align}
&O(ap^2\bm{m})&
&\vev{\S\hata^\dag + \hata\S^\dag}^2\vev{D_\m\S D_\m\S^\dag},
\vev{\hata\hata^\dag}\vev{D_\m\S D_\m\S^\dag},
\vev{D_\m\S\hata^\dag + \hata D_\m\S^\dag}^2,\notag\\
&&&\underbrace{\vev{D_\m\S\hata^\dag D_\m\S\hata^\dag
+ D_\m\S^\dag \hata D_\m\S^\dag\hata}}_{S_4},
\underbrace{\vev{D^2\S\hata^\dag + \hata D^2\S^\dag}
\vev{\S\hata^\dag + \hata \S^\dag}}_{S_5},\\
&O(a\bm{m}^2)&
&\vev{\S\hata^\dag + \hata\S^\dag}^2\vev{\S \tchi^\dag + \tchi\S^\dag},
\vev{\hata \hata^\dag}\vev{\S \tchi^\dag + \tchi\S^\dag},
\vev{\hata\tchi^\dag + \tchi\hata^\dag}\vev{\S\hata^\dag + \hata\S^\dag},
\end{align}
where $S_4, S_5$ terms vanish in the untwisted theory.
In the $l_\m = r_\m = 0$ limit, they become
\begin{align}
S_4 &\to \frac{1}{2}
\vev{\p_\m\S\hata^\dag + \hata \p_\m\S^\dag}
\vev{\p_\m\S\hata^\dag + \hata \p_\m\S^\dag}
- \vev{\hata\hata^\dag}\vev{\p_\m\S \p_\m\S^\dag},\\
S_5 &\to - \vev{\p_\m\S\hata^\dag + \hata \p_\m\S^\dag}^2.
\end{align}

\subsection{Renormalization for a vacuum expectation value}

Since there exist three point functions
in the Lagrangian (\ref{eq:pi3Lagrangian}),
we have one loop diagrams for a vacuum expectation value.
We obtain
\begin{align}
 \parbox[c]{20mm}{\includegraphics{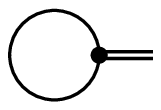}} &=
\frac{a 2B_0m' (2c_0 - 2\tilde c_2)\sin\phi_0 + 2c_2 a^2 \sin 2\phi_0}{2f_0}
 I_\pm,\\
 \parbox[c]{20mm}{\includegraphics{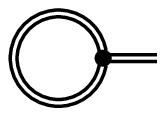}} &=
\frac{a 2B_0m' (c_0 - 3\tilde c_2)\sin\phi_0 + 3c_2 a^2 \sin 2\phi_0}{2f_0}
 I_0,
\end{align}
where  the single(double) line represents the charged(neutral)
pion, $I_{\pm/0}$ denotes a contribution from a charged or neutral pion loop, which is given by
\begin{align}
I_{\pm/0} &= \int\frac{d^4p}{(2\pi)^2}\frac{1}{p^2 + (m_\pi^{\pm/0})^2}\notag\\
&= \frac{(m_\pi^{\pm/0})^2}{16\pi^2}\left[- \frac{2}{\epsilon} + \g_E - 1
+ \log\frac{(m_\pi^{\pm/0})^2}{4\pi} \right].
\end{align}
where $\epsilon$ is the dimensional regulator and $\gamma_E$ is the
Euler-Mascheroni constant.
One-loop effects to the vacuum expectation value are summarized as
\begin{align}
 \lag_\Loop^{(1)} &= \frac{1}{f_0}(A_\pm I_\pm + A_0 I_0)\pi_3,
 \\
 A_\pm &= c_2 a^2 \sin 2\phi_0 + (c_0 - \tilde c_2) a (2B_0m') \sin\phi_0,
\\
 A_0 &= \frac{3}{2}c_2 a^2 \sin 2\phi_0 + \frac{1}{2}(c_0 - 3\tilde c_2)
 a (2B_0m') \sin\phi_0.
\end{align}

Using the gap equation (\ref{eq:WChPT_gap_eq2}), on the other hand,
contributions from $\lag_{\NLO+\NSLO}$ are given by
\begin{align}
\lag_{\NLO+\NSLO}^{(1)} &= -\frac{8}{f_0}\bigl[
a (2B_0m')^2(8V_4 + V_5 + 4 \tilde c_2 L_{68})\sin\phi_0\notag\\
&+a^2 (2B_0m')(8X_4 + X_6 - 4c_2L_{68})\sin 2\phi_0\bigr]\pi_3.
\end{align}
In order to cancel divergences from the 1-loop integral,
low energy constants in $\lag_{\NLO+\NSLO}$ need to be renormalized,
and renormalized low energy constants are given by%
\begin{align}
L_i &= L^r_i(\m_\ChPT) + \frac{l_i}{32\pi^2}R, \\
V'_i &= V^r_i(\m_\ChPT) + \frac{v_i}{32\pi^2}R, \\
X'_i &= X^r_i(\m_\ChPT) + \frac{x_i}{32\pi^2}R,
\end{align}
where the argument $\m_\ChPT$ is a renormalization scale
and $R$ is  defined by
\begin{align}
 R &= - \frac{2}{\epsilon} - \log(4\pi) + \gamma_E - 1,
\end{align}
which cancels the 1-loop divergence in dimensional regularization.
To renormalize the vacuum expectation value correctly,
we need renormalization conditions,  eqs. (\ref{eq:rcondition_v1}) and (\ref{eq:rcondition_v2}) ,
and the renormalized Lagrangian for $\pi$ is written as
\begin{align}
 \lag_R^{(1)} &= \left.\frac{\p}{\p \phi}
\lag_{\LO + \SLO}^{(1)}\right|_{\phi = \phi_0} \Delta \phi
+ \lag_\Loop^{(1)} + \lag_{\NLO+\NSLO}^{(1)}\notag \\
&= \left[(m_\pi^0)^2 f_0 \Delta \phi + f_0(A_\pm L_\pm + A_0L_0)
- \frac{8}{f_0}\left\{a(2B_0m')^2C_1 + a^2(2B_0m')C_2\right\}
\right]\pi_3,
\end{align}
where the low energy coefficients $C_1$, $C_2$ are given as,
\begin{align}
 C_1 &= (8V^r_4 + V^r_5 + 4 \tilde c_2 L^r_{68}) \sin\phi_0,\\
 C_2 &= (8X^r_4 + X^r_6 - 4c_2L^r_{68}) \sin 2\phi_0.
\end{align}

\subsection{Renormalization for mass}
The 1-loop diagrams, which contribute to the pion mass term,
can be classified into two types.
The first type are made from a four-point vertex,
while the second ones are made from two three-point vertices.
As the first type, we have following diagrams.
\begin{center}
\includegraphics{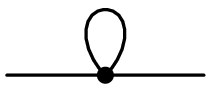}
\includegraphics{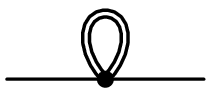}
\includegraphics{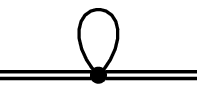}
\includegraphics{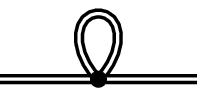}
\end{center}
In the second type, there are three diagrams.
\begin{center}
\includegraphics{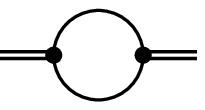}
\includegraphics{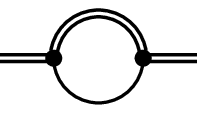}
\includegraphics{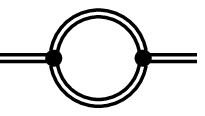}
\end{center}
In the same way as tadpole diagrams,
the contributions of the first type diagrams are given by
\begin{align}
\parbox[c]{25mm}{\includegraphics{f_diagram/prop_c_loop_c.eps}}
&= \frac{1}{2f_0^2}\left[ Z^\pm_\pm (\p_\m\pi_\pm)^2
+ (\Delta M^\pm_\pm)^2 \pi_\pm^2
\right]I_\pm,\label{eq:mass_loop1}\\
\parbox[c]{25mm}{\includegraphics{f_diagram/prop_c_loop_n.eps}}
&= \frac{1}{2f_0^2}\left[ Z^\pm_0 (\p_\m\pi_\pm)^2
+ (\Delta M^\pm_0)^2 \pi_\pm^2
\right]I_0,\label{eq:mass_loop2}\\
\parbox[c]{25mm}{\includegraphics{f_diagram/prop_n_loop_c.eps}}
&= \frac{1}{2f_0^2}\left[ Z^0_\pm (\p_\m\pi_0)^2
+ (\Delta M^0_\pm)^2 \pi_0^2
\right]I_\pm,\label{eq:mass_loop3}\\
\parbox[c]{25mm}{\includegraphics{f_diagram/prop_n_loop_n.eps}}
&= \frac{1}{2f_0^2}\left[ Z^0_0 (\p_\m\pi_0)^2
+ (\Delta M^0_0)^2 \pi_0^2
\right]I_0.\label{eq:mass_loop4}
\end{align}
where each coefficient is given by
\begin{align}
 Z^\pm_\pm &= -\frac{1}{3} - c_0a\cos\phi_0,&
 (\Delta M^\pm_\pm)^2 &= m_\pi^2 \left\{
-\frac{1}{3} + a(c_0 - 4\tilde c_2)\cos\phi_0
\right\} + 4c_2 a^2\cos^2\phi_0,\\
 Z^\pm_0 &= -\frac{1}{3} - \frac{c_0a\cos\phi_0}{2},&
 (\Delta M^\pm_0)^2 &= m_\pi^2 \left\{
\frac{1}{6} + \frac{a(c_0 - 2\tilde c_2)\cos\phi_0}{2}
\right\} + c_2 a^2\cos^2\phi_0,\\
 Z^0_\pm &= -\frac{2}{3} - c_0a\cos\phi_0,&
 (\Delta M^0_\pm)^2 &= m_\pi^2 \left\{
\frac{1}{3} + a(c_0 - 2\tilde c_2)\cos\phi_0
\right\} + c_2a^2(2\cos^2\phi_0 - \frac{4}{3}\sin^2\phi_0),\\
 Z^0_0 &= -\frac{c_0a\cos\phi_0}{2},&
 (\Delta M^0_0)^2 &= m_\pi^2 \left\{
-\frac{1}{2} + \frac{a(c_0 - 6\tilde c_2)\cos\phi_0}{2}
\right\} + c_2 a^2(3\cos^2\phi_0 - 4\sin^2\phi_0).
\end{align}
On the other hand,  the second type diagrams give $\SLO^2$ effects as
\begin{align}
\parbox[c]{25mm}{\includegraphics{f_diagram/sun_cc.eps}} &=
(\SLO)^2\int d^4p \frac{1}{p^2+m_\pi^2}\frac{1}{p^2+m_\pi^2}\notag\\
&= \SLO^2.
\end{align}
Therefore the second type diagrams are higher order than NLO
and we do not consider them in this paper.
We then obtain the renormalized Lagrangian for $\pi^2$ at 1-loop as
\begin{align}
 \lag_R^{(2)} &= \lag_{\LO + \SLO}^{(2)}
+ \lag_\Loop^{(2)} + \lag_{\NLO +\NSLO}^{(2)},\notag\\
&= \frac{1}{2} \left\{
1 + \frac{Z^a_bI_b}{f_0^2} + C_\NLO^{2p,a}
\right\} (\p_\mu\pi_a)^2 + \frac{1}{2} \left\{
(m_\pi^a)^2 + \frac{M^a_bI_b}{f_0^2} + C_\NLO^{2,a}
\right\} \pi^2.
\end{align}
From this result,  the renormalized mass is given by
\begin{align}
 (m_\pi^a)_\NLO^2 &= \left\{
(m_\pi^a)^2 + \frac{M^a_bI_b}{f_0^2} + C_\NLO^{2,a}
\right\} \left\{
1 + \frac{Z^a_bI_b}{f_0^2} + C_\NLO^{2p,a}
\right\}^{-1},\notag\\
&= (m_\pi^a)^2 \left\{1 - \frac{Z^a_bI_b}{f_0^2} - C_\NLO^{2p,a}\right\}
+ \frac{M^a_bI_b}{f_0^2} + C_\NLO^{2,a}\notag\\
&= \left.(m_\pi^a)^2_\LO\right|_{\phi_0 \to \phi_0 + \D\phi}\notag\\
& + \sum_{b = \pm, 0}\left( m_\pi^2 B^a_b + a^2 Q^a_b \right)L_b\notag \\
&-\frac{16}{f_0^2} \left\{E_1^a (2B_0m')^2 + E_2^a a(2B_0m')^2+ E_3^a a^2(2B_0m')\right\},
\end{align}
where coefficients of the chiral log terms are given in eqs. (\ref{eq:constA})--(\ref{eq:constD}),
while
\begin{align}
E_1^a &= L^r_{45} - 2L^r_{68},\\
E_2^a &= (12V'^r_4 + V^r_5/2 + V^r_6 + 4V^r_1 + V^r_2 - \tV^r_{23}/2
+ 2\tilde c_2 L^r_{45})\cos\phi_0,\\
E_3^a &= (6X^r_4 + X^r_5/2 + X^r_6/2 + 2X^r_1 + X^r_2/2
- \tX^r_{23}/2 - c_2L^r_{45})\notag\\
&\qquad + \cos 2\phi_0(6X^r_4 + X^r_6/2 + 2X^r_1 - c_2L^r_{45})\notag\\
&\qquad - \d_{a0}(4X^r_4 - X^r_3 - \tX^r_{23}/2 + \tX^r_3 - c_2L^r_{45})
(1 - \cos 2\phi_0)
\end{align}
for NLO low energy constants.

\subsection{Renormalization for decay constant}

Since we have $O(\pi^3)$ terms (\ref{eq:A12_pi3}, \ref{eq:A3_pi3})
in the axial current, the 1-loop contribution to the decay constant of the charged pion  is given by
\begin{align}
\vev{0|A_\m^a|\pi_a(p)}_{\Loop} &= - f_0p_\m C_\Loop^{f,a},\notag\\
C_\Loop^{f,a} &= \left[
\left( \frac{2}{3f_0^2} + \frac{c_0a\cos\phi_0}{2f_0^2} \right) \vev{\pi_b\pi_b}
- \frac{2}{3f_0^2} \vev{\pi_a\pi_a} \right] \cos\phi_0
-\frac{c_0a\sin^2\phi_0}{f_0^2}\vev{\pi_3\pi_3}\notag\\
&=  \left[
\left( \frac{2}{3} + c_0a\cos\phi_0 \right) L_\pm
+ \left( \frac{2}{3} + \frac{3c_0a\cos\phi_0}{2} \right) L_0 \right] \cos\phi_0
-c_0aL_0.
\end{align}
For the neutral pion,
\begin{align}
\vev{0|A_\m^3|\pi_3(p)}_{\Loop} &= - f_0p_\m C_\Loop^{f,3},\notag\\
C_\Loop^{f,3} &= \left( \frac{2}{3f_0^2} + \frac{c_0a\cos\phi_0}{2f_0^2} \right) \vev{\pi_b\pi_b}
- \frac{2}{3f_0^2} \vev{\pi_3\pi_3}\notag\\
&= \left( \frac{4}{3} + c_0a\cos\phi_0 \right) L_\pm
+ \frac{c_0a\cos\phi_0}{2} L_0.
\end{align}
Therefore, we obtain the renormalized decay constant
\begin{align}
\vev{0|A_\m^a|\pi_a(p)}|_R = \left\{
1 + \frac{Z^a_bI_b}{f_0^2} + C_\NLO^{2p,a}
\right\}^{-1/2}\vev{0|A_\m^a|\pi_a(p)}
+ \vev{0|A_\m^a|\pi_a(p)}_{\Loop} + \vev{0|A_\m^a|\pi_a(p)}_{\NLO},
\end{align}
where the first term is a contribution for the renormalization of the pion field,
and the second and third terms represent the 1-loop and the ``NLO'' contributions, respectively.
In the same way as the vacuum expectation value and the pion mass,
we obtain the decay constant at NLO as
\begin{align}
f_{PS,\NLO}^a &= \left.f_{PS,\LO}^a\right|_{\phi = \phi_0 + \D\phi}
+ f^a_{PS,\LO}\Bigl[
\sum_{b = \pm,0} F_b^a L_b\notag\\
& \qquad+ \frac{16}{f_0^2} \left\{
(2B_0m') H_1^a + a(2B_0m') H_2^a + a^2 H_3^a\right\}\Bigr]\notag\\
& \qquad + \D f_\pm (1 - \d_{3a}).
\label{eq:decay_const_app}
\end{align}
where coefficients for the chiral log terms are given in eqs. (\ref{eq:const2}) and (\ref{eq:const3}),
while
\begin{align}
H_1^a &= L^r_{45}/2,\\
H_2^a &= 2V^r_1 + V^r_2/2 - \tV^r_{23}/4 - 2 \tV^r_{3m} -2 \tV^r_{3a}\d_{3a},\\
H_3^a &= X^r_1 + X^r_2/4 - \tX^r_{23}/4 - 2 \tX^r_3 + X^r_1 \cos 2\phi_0\notag\\
&\qquad + (X^r_3/2 + \tX^r_{23}/4 + 3\tX^r_3/2)(1 - \cos 2\phi)\d_{3a}
\end{align}
for  NLO low energy constants.
Note here that an additive renormalization term
for the decay constant of the charged pion exists:
\begin{align}
\D f_\pm &= - f_0 \left[-a (2B_0m') 2V^r_{3a} + c_0a L_0.
\right]
\end{align}

We finally consider the PCAC mass, defined by
\begin{align}
m_\PCAC &=
\frac{\sum_{\bm{x}}\int\frac{d^3p}{(2\pi)^3}
\vev{ 0|\p_4 A_4^a({\bm{x}},t) | \pi(p)}\frac{1}{2E_p}\vev{\pi(p)|P^a(0)|0}}
{2\sum_{\bm{x}} \int\frac{d^3p}{(2\pi)^3}
\vev{0| P^a({\bm{x}},t) | \pi(p)}\frac{1}{2E_p}\vev{\pi(p)|P^a(0)|0}} \\
&= \frac{f_{PS} m_\pi^2}{2Z^a_{\rm PS}},
\end{align}
Using the same spurion analysis as for the axial currents,
the pseudo scalar density and its renormalization factor $Z^a_{\rm PS}$ are given by
\begin{align}
P^a &= if_0 B_0 (1 + \tilde c_2 a \cos\phi_0)\pi_a \quad (a = 1,2),\\
Z^a_{\rm PS} &= f_0 B_0 (1 + \tilde c_2 a \cos\phi_0)
\end{align}
at the tree-level.
Therefore
the LO PCAC quark mass becomes
\begin{equation}
m_\PCAC = \frac{m_\pi^2 \cos\phi_0}{2B_0(1 + \tilde c_2 \cos\phi_0)}.
\end{equation}

\end{document}